\documentclass[12pt]{article}
\usepackage{graphicx}
\usepackage{setspace}
\usepackage{rotating}
\usepackage{natbib}
\usepackage{amsmath, amssymb, enumerate, amsthm}
\usepackage{dsfont}
\usepackage{lscape,amssymb, amsmath,verbatim,graphicx}
\usepackage{epsfig,float}
\usepackage{graphicx}
\usepackage{epsfig}
\usepackage{booktabs}

\newcommand{\sa}{\begin{align*}
}
\newcommand{\se}{\end{align*}
}
\numberwithin{table}{section}
\numberwithin{figure}{section}

\usepackage{subfigure}

\usepackage[papersize={8.5in,11in},left=0.8in,right=0.8in,top=0.9in,bottom=0.9in]{geometry}

\allowdisplaybreaks
\def\beq{\begin{equation}}
\def\eeq{\end{equation}}

\numberwithin{equation}{section}
\numberwithin{theorem}{section}

\usepackage{hyperref}
\hypersetup{
	colorlinks=true,
	linkcolor=blue,
	citecolor=blue,
	filecolor=magenta,      
	urlcolor=cyan,
}
\title{\textbf{Functional Boxplots for Outlier Detection in Additive Manufacturing}}
\small{\author{Ahmad Mozaffari \\
  \makeatother
  {\small \it  Department of Statistics and Actuarial Science, University of Waterloo, Canada} \\
 Shojaeddin Chenouri \\
  {\small \it Department of Statistics and Actuarial Science, University of Waterloo, Canada} \\
  Ehsan Toyserkani \\
    {\small \it  Department of Mechanical and Mechatronics Engineering, University of Waterloo, Canada} \\
     Usman Ali \\
    {\small \it  King Fahd University of Petroleum and Minerals, Saudi Arabia} 
}
\date{}
\begin{document}
\maketitle
\pagestyle{myheadings}
\begin{abstract}
Additive manufacturing (AM), also known as 3D printing, is one of the most promising digital manufacturing technologies, thanks to its potential to produce highly complex geometries rapidly. AM has been promoted from a prototyping methodology to a serial production platform for which precise process monitoring and control strategies to guarantee the accuracy of products are required. This need has motivated practitioners to focus on designing process monitoring tools to improve the accuracy of produced geometries. In line with the emerging interest, in the current investigation, a novel strategy is proposed which uses functional representation of in-plane contours to come up with statistical boxplots with the goal of detecting outlying AM products. The method can be used for process monitoring during AM production to automatically detect defective products in an online fashion. To ensure the considered method has an acceptable potential, different complex 3D geometries are considered and undergo different types of stochastic perturbations to collect data for outlier detection. The results of the conducted simulation are very promising and reveal the reliability of the proposed method for detecting products with statistically significant deformation.

\end{abstract}

\maketitle

\section{Introduction}

Additive manufacturing (AM), also called 3D printing, is the process of fabricating 3D products,  layer-by-layer under the control of computers. This is done using a collection of digital data (or, if available, a mathematical model) representing the geometry of interest which enables computer aided design (CAD). Unlike conventional manufacturing technologies that involve material removal in their process, AM produces objects layer-by-layer in an additive manner, and thus, there is no waste of material \citep{tapia2014}. This can benefit industrialists economically. Also, it can prevent life hazards in cases that the material are poisonous, because of no waste during the production \citep{qianch2017}. Such a feature enables AM to directly produce products with complicated 3D geometries. This makes AM a game changing technology in the manufacturing industry and over the past few decades, this technology has gained a tremendous interest with a wide range of applications, including architecture design \citep{yiquzhli2018}, biomedicine \citep{melfegr2010}, aeronautics and astronautics \citep{schil2015}, and \textit{etc}. Comprehensive reviews on the applications of AM to different key industries and its societal impact can be found in \cite{gule2013, hulimoho2013}. 

However, in spite of its obvious advantages, todays AM technology suffers from two important flaws, namely unacceptable quality and repeatability of products. This incurs some technical difficulties (especially for applications like aerospace in which product quality is of high importance), and has enforced researchers to seek for methods and techniques to overcome the mentioned problems. Fortunately, AM is performed under full control of computers, and this enables the use of process controlling and monitoring computer programs to guarantee the quality and repeatability of desired products. In line with such interests, in recent years, the use of statistical monitoring and outlier detection concepts has been investigated by researchers and is continued to be considered as one of the most important research goals of additive manufacturers \citep{tapia2014, cohudats2018}. 

The purpose of the current investigation is to probe the potential of some recent statistical tools and concepts to be used for quality assurance in material extrusion AM method. In particular, an outlier detection algorithm is implemented for detecting products which deviate significantly from desired geometry. In general, outlier detection in AM is a complicated task, since data presenting geometries of interest usually have functional nature (in the form of planar or spatial). Therefore, more care should be exerted to make sure the considered outlier detection methods are capable of identifying the sources of variability as well as significant outliers in a set of collected functional data \citep{srikla2016}. The adopted statistical method decomposes a set of functionals into three different modules, i.e. translation, amplitude, and phase, and creates distinct geometrical boxplots for each of them to identify outliers \citep{xiekubhas2017}. This method can also be used for process monitoring in AM, since after collecting a set of geometrical data, it is possible to design boxplots which can be later used in the process to automatically detect outlying products (diagnosis of possible faults).

From statistical viewpoint, the salient asset of the considered method is that it does not neglect any source of variation in the functional data, and thus, can be viewed as a reliable outlier detection method compared to projection based methods, e.g. functional principle component analysis (FPCA), which can identify outliers only in certain directions. Also, from the practical viewpoint, this method is advantageous, since as a boxplot outlier detection technique, it provides a graphical tool which can be easily interpreted by practitioners who do not necessarily have a firm statistical background. It is worth pointing out that to have an effective outlier detection method, it is important to make sure the geometry of interest can be modelled efficiently, which then can be used for further processes. In the next section, a review on the modelling and monitoring tools available in the literature for AM will be presented for the better understanding of the problem at hand. 

The primary goal of the current simulation study is to evaluate the performance of the considered method for detecting significantly deformed 3D geometries produced during a 3D printing process. To do so, the geometries of interest are mathematically modelled in the form of multivariate (spatial) functionals and statistical shape analysis methods are used to identify outliers.

The rest of the paper is organized as follows. A review on the existing modelling and process monitoring techniques used in AM is presented in Section \ref{litreview}. Section \ref{method} is devoted to the implementation of the considered method. Also, some of the properties of the method are discussed in this section. In Section \ref{3DModel}, the mathematical formulation of a benchmark 3D geometry as well as different real datasets of 3D printed products are presented. Section \ref{MCMCsetup} explains the Monte Carlo (MC) simulation performed to collect different databases for the statistical analysis. Section \ref{simresults} presents the simulation results, as well as reports regarding the performance of the proposed method for the considered cases. Section \ref{discussions} provides some general discussions about the potentials of the considered approach as well as some clues on how to effectively use them in practice.  Finally, the concluding remarks together with some outlines for future investigation are given in Section \ref{conclusions}.

\section{Literature Review} \label{litreview}

As mentioned, the conducted research on using statistical methods for AM comprises of developing methods for modelling geometrical objects and process monitoring / control. The details of the conducted research and their findings are given in two separate parts. In the first part, the focus is on explaining the methods useful for modelling geometrical objects, while the second part discusses the methods advantageous for process monitoring / control in AM. It should be noted that research on both modelling and process monitoring / control in AM is rather recent, and there is a need for further investigations to fill in the gap. Hence, in what follows this section, some of the main conducted research are discussed to clarify the current state-of-the-art, and also to show what needs to be done to further enrich the literature. 

\subsection{Modelling Geometrical Objects}
The methods proposed so far for AM applications includes both numerical and analytical models. In numerical approaches, the goal is to use a predefined grid / mesh to scan the surface of the object, while in analytical models one intends to use geometrical tools to come up with an analytical model for the shapes under study. Research on using numerical and analytical approaches for modelling shapes to enable process monitoring / control have been pursued in tandem. In \cite{sadahuzh2014}, an analytical framework based on Fourier basis functions was implemented to model 3D objects, which was then used for detecting and modelling interference, with the goal of better understanding quality control for AM. In \cite{zhan2015}, a geometrical approach called surface-based modification algorithm (SMA) was proposed to adaptively and locally increase the density of triangular facets of Stereolithography (STL) approximation of CAD models. Simulation indicated improvement in the part error of SMA-enhanced STL models compared to original SLT. In \cite{huang2016}, an analytical concept based on minimum area deviation (MAD) and minimum volume deviation (MVD) for both in-plane (2D) and out-of-plane (3D) modelling of 3D objects were proposed with the goal of optimal compensation of deformations in AM. The model enabled both layer-by-layer and entire shape analysis of products. By applying the proposed approach to different geometries, the acceptable potential of the proposed approach was proven. In \cite{chtswa2017}, the concept of transfer learning was used to model shape deviation in AM. The key idea behind using transfer learning was that it is almost impossible to develop a unique model for all of the 3D objects produced in AM. The modelling error was decomposed to shape-dependent and shape-independent errors. The simulation indicated the acceptable performance of the proposed concept. In \cite{zhanma2017}, a deviation modelling concept was used based on STL. A novel transformation method was proposed based on contour point displacement, for which, in each slice, systematic deviations were
represented by polar and radial functions while random deviations were captured by translating the contour points using a distance derived from the random field theory. Performed simulation indicated that the method acceptably predicted both repeatable and unexpected deviations. In \cite{wasohuts2017}, a hybrid model was proposed for in-plane shape deviation modelling and compensation in fused deposition modelling (FDM) processes. The hybrid algorithm decomposed the errors into two different major sources, (1) the positioning error of extruder for which a Kriging model was used, and (2) shape deformation due to process error for which a Fourier basis function was taken into account. Simulation results demonstrated that the proposed hybrid model could accurately manage different error sources in AM. In \cite{luhu2017}, for the first time, the idea of prescriptive-predictive modelling and compensation was proposed for the in-plane deformation analysis of 3D freeform products. The prescriptive modelling approach was used to enable analysis beyond the experimental scope of test shapes (deformation analysis for untried categories of shapes), while conventional predictive modelling did the prediction with the experimental category. In \cite{zhkeanma2017}, a review of shape deviation modelling for AM was conducted, and it was stated that skin model shape paradigm could be a promising method for the analysis of shape deviations of products. 

\subsection{Process Monitoring / Control in AM} The other important issue which can be considered as the final goal for the problem at hand is the design of efficient process monitoring / controlling algorithms. Due to its importance, different researchers have done remarkable efforts to design effective methods for improving the quality of products in AM. In \cite{coli2010}, a feedback control algorithm was proposed for monitoring the quality of geometrical parts in solid freeform fabrication (SFF) process. Conducted simulation indicated that the proposed method could compensate a broader range of geometrical errors compared to other methods, and also was computationally inexpensive which made it proper for real-time monitoring application. In \cite{bebelikr2010}, different aspects and parameters affecting the quality control of products in AM were discussed. In \cite{tapia2014}, a review on the advances in process monitoring and control for metal-based AM was performed, which can be viewed as the first comprehensive review on the subject. Based on a throughout analysis, some remarks for future researches were pointed out. In \cite{hunoxuchsoda2014}, a statistical predictive modelling was proposed to compensate the geometrical deviations of three-dimensional printed products. Simulation on a number of polyhedron shapes demonstrated the capability of the method for predicting and compensating errors for a wide class of products. In \cite{lumi2015}, a layer-to-layer predictive modelling and feedback control was proposed to control ink-jet 3D printing process. Also, an extension of the algorithm was implemented for two material printing process to enable the printing of complex 3D geometries. Simulation results indicated that the method could reduce the shrinking of parts, and made top layer surface smoother. In \cite{huzhsada2015}, an analytical framework was proposed for the offline optimal compensation of shape shrinkage for 3D printing processes. Experiments indicated that the proposed method can reduce the geometrical error of cylindrical products by an order of magnitude. In \cite{ridwan2015}, the potential of the process monitoring of AM with the goal of improving the quality of products was investigated. In \cite{jiqihu2016}, an out-of-plane predictive control algorithm was proposed for the offline compensation of geometric errors in AM. Simulation results indicated the accuracy of the proposed control approach. In \cite{evhistlecl2016}, a review on the in-situ process monitoring and metrological methods for metal AM was done. In this review, the suitability of inspection methodologies compatible with AM for identifying typical material discontinuities and failure modes were investigated, and based on that, some remarks for future researches were given. In \cite{xukwzhch2017}, a reverse compensation framework for controlling of shape deformation in AM was proposed. Simulations indicated improvements in the compensation of geometrical errors. In \cite{chahmo2017}, a review on process monitoring systems for metal AM was performed, and a real-time inspection method together with a closed-loop control algorithm was proposed for quality control in AM. In \cite{colosimo2018}, a review on modeling and monitoring methods for spatial and image data with application to AM was carried out, and some of the main challenges were discussed. Also solutions to these problems together with some open issues for future research were outlined. In \cite{grdeprco2018}, a process monitoring method that consists of a feature extraction technique and a data-driven statistical rule was proposed for monitoring powder bed fusion processes. Comparative analysis demonstrated the effectiveness of the proposed method. In \cite{zhqi2018}, a phase I control chart was proposed for monitoring of spatial surface data from 3D printing. A comparative study was conducted to prove the efficacy of the proposed control chart. In \cite{zhqi2018b}, a non-parametric control chart was proposed for phase II monitoring of spatial surfaces produced in 3D printing. The veracity of the proposed method was ascertained by numerical simulation. As can be inferred, researchers are becoming more and more interested in using standard statistical quality control methods for AM, and it is expected that in near future, more specific investigations will be conducted to fill the gap in the current literature.
 
 \section{Implementation of the Outlier Detection Method} \label{method}
 
The considered method is an extension of boxplots used for outlier detection when the collected data are functionals. Boxplots  are non-parametric tools that are used to study the variation in a dataset regardless of their underlying distribution. Upon proper design (taking the starting and endpoint of whiskers as desired cut-off values), boxplots can be used for outlier detection. Boxplots are global outlier detection methods which use the entire sample set to determine the cut-off values. One of the important features of boxplots over other outlier detection concepts (such as distance-based, density-based, and deviation-based methods) is their simple statistical structure, which make them easy to compute and interpret. Also, since boxplots are graphical tools, the outlyingness of a given point can be clearly determined from the plot, even by those who do not have statistical background. This interesting property makes boxplots a suitable outlier detection tool, especially for AM applications, as practitioners, who may not have statistical background, need to do the process monitoring in real-time.

\begin{figure} 
 	\begin{center}
 		\includegraphics[width=0.8\textwidth,angle =0, scale=1]{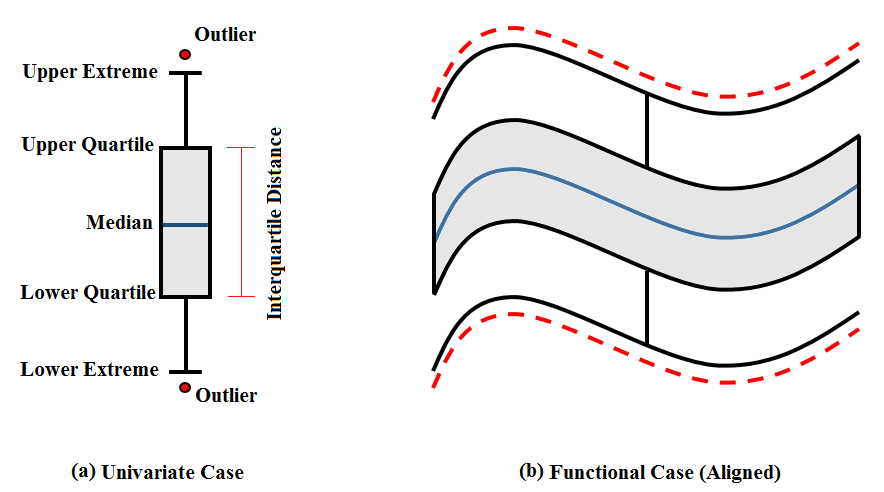}
 	\end{center} \caption{Schematic illustration of (a) univariate boxplot and (b) its corresponding functional boxplot (functions are aligned and belong to the same equivalent phase class).}\label{FcnBoxplot}
 \end{figure}

As mentioned, to make boxplots usable for the current application, some extensions need to be done. This is because the collected data representing a geometrical shape usually has a functional nature. Fortunately, functional extensions of boxplot are available in the literature \citep{xiekubhas2017, suge2011}. Figure \ref{FcnBoxplot} depicts a univariate boxplot and the corresponding functional boxplot under the assumption that functionals are registered. It can be seen that these two versions of boxplots share many similarities in their structures. It should be noted that the functional boxplot shown in the figure is obtained under the optimistic assumption that functions belong to the equivalent phase class. However, in practice, there is also a variation in the phase of the collected functional data, which makes the shape of resulting boxplot complicated. One way to cope with this difficulty is to decompose the sources of variation of functaionals into the variation of phase, amplitude, and (if necessary) translation components \citep{xiekubhas2017}. By doing so, each source of variation is analyzed in a separate space, and independent boxplots are developed to study the variation in the amplitude, phase and translation components of functional observations. Such a functional outlier detection analysis has some advantages over projection-based and point-wise boxplots:

\begin{itemize}
	\item Projection-based approaches, e.g. functional principal component
	analysis (FPCA), can identify outliers only in certain directions while the considered method does the analysis in the original functional space, and thus, the results are more reliable.	
	\item Point-wise boxplots may not identify sources of variation properly
	due to discretization, as well as ignoring within curve dependencies
	(e.g. temporal or spatial dependencies) while the considered approach does not suffer from such flaws.
\end{itemize}

The mathematical tools and definitions as well as the detailed steps for developing amplitude and phase boxplots are given in the following subsections.

\subsection{Mathematical tools} \label{setnotations}

Throuought the analysis, it is assumed that the functions are defined in the range [0,1]. Let's represent the space of functions as $$\mathcal{F} = \{ f: [0,1] \to \mathbb{R} \mid f\ \text{is\ absolutely\ continuous} \}\,.$$ Due to absolute continuity, $f(t) = f(0) + \int_{0}^{t} \dot{f}(s) ds$, where $\dot{f}$ is the derivative of the function $f$. Also, let's define the space of warping functions as $\Gamma = \{ \gamma:[0,1] \to [0,1] \mid \dot{\gamma}>0,\ \gamma(0)=0,\ \gamma(1)=1 \}$. The warping functions $\gamma$ are orientation-preserving diffeomorphisms. For any given function $f$ and warping function $\gamma$, the warping of $f$ using $\gamma$ is presented by decomposition $f \circ \gamma$. The main idea is that warping function represents the phase variability and should be analyzed separately from amplitude. Since the original function space $\mathcal{F}$ is not invariant under $L^2$ (i.e. $||f_1 - f_2||_{_2} \neq || f_1 \circ \gamma - f_2 \circ \gamma||_{_2}$), a bijection mapping called square root slope function (SRSF),$\mathcal{Q}: \mathcal{F} \to L^2\big([0,1],\mathbb{R}\big)$, is used, which is mathematically defined as $Q(f) = q = sgn(\dot{f}) \sqrt{| \dot{f} |}$. By definition, functions $q$ are invariant under translation, and also they act under $L^2$ by isometry $||q_1 - q_2||_{_2} = || q_1 \circ \gamma - q_2 \circ \gamma||_{_2}$. This feature faciliates the calculation of optimal warping for registeration, which is necessary for the decomposition of phase and amplitude components. After decomposition, the elements of SRSF space $L^2\big([0,1],\mathbb{R}\big)$ are represented by their corresponding elements in the amplitude space $L^2/\Gamma = \{[q] | q \in L^2 \}$, and their respective warping functions in phase space $\Gamma$. Note that $[q]$ represents the equivalent class $[q] = \{q \circ \gamma | \gamma \in \Gamma \}$. Since there is a bjection relation between $\mathcal{F}$ and SRSF space, each $q$ corresponds to a unique function $f$ in the original function space $\mathcal{F}$. Also, it can be shown that $q \circ \gamma = (q,\gamma) \sqrt{\dot{\gamma}}$. It is obvious that development of boxplots in amplitude and phase spaces is more challenging than the standard boxplots designed in $\mathbb{R}$, and requires geometrical tools from the corresponding spaces. These issues will be studied in details in what follows this section.

\subsection{Amplitude Boxplot}

Just like standard boxplot, the design of amplitude boxplot requires information like median, first and third quartiles, and cut-off functions. To obtain those information, the geometry of the amplitude space $\mathcal{F}/\Gamma$ (or equivalently $L_2/\Gamma$) should be taken into account. Assume that a set of functional data $\{f_1,f_2,\cdots,f_N\}$ are available. The median can be computed as:
\begin{equation} \label{D_a}
[\bar{q}] = \arg \min_{[q] \in L^2 / \Gamma} \sum_{i=1}^{N} D_a ([q_i],[q]) \ ,
\end{equation}
where $D_a ([q_1],[q_2]) = \inf_{\gamma \in \Gamma} || q_1 - (q_2 , \gamma) ||_{_2}$. Solution can be found using gradient descent algorithm. After finding the median $[\bar{q}]$ (and corresponding $[\bar{f}]$), select an element from this equivalent class (denoted by $\bar{q}$) using orbit centering method \citep{srikla2016}. Align all of the functions to the median using $D_a$. This gives us the set of aligned functions $\{\tilde{f}_1, \tilde{f}_2, ..., \tilde{f}_N \}$, aligned SRSFs $\{\tilde{q}_1, \tilde{q}_2, ..., \tilde{q}_N \}$, amplitude distances $\{D_a^1, D_a^2, ..., D_a^N \}$, and optimal phase functions $\{\gamma_1, \gamma_2, ..., \gamma_N \}$. The obtained information can be used to find the central region of the boxplot. To do so, one needs to order the amplitude components $\{\tilde{q}_{_1}, \tilde{q}_{_2}, ..., \tilde{q}_{_N} \}$ based on the calculated distances (which show their proximity to median), and extract 50\% of amplitude functions that are closest to median. Let's denote this order set by $\{ \tilde{q}_{_{(1)}}, \tilde{q}_{_{(2)}}, ..., \tilde{q}_{_{( \lfloor N/2 \rfloor)}} \}$. Then, the first and third quartiles, denoted by $\tilde{q}_{_{Q_1}}$ and $\tilde{q}_{_{Q_3}}$, can be obtained as \citep{xiekubhas2017}:
\small
\[
(\tilde{q}_{Q_1} , \tilde{q}_{Q_3}) = \underset{ q_1 , q_2 \in \{ \tilde{q}_{(1)}, \tilde{q}_{(2)}, ..., \tilde{q}_{( \lfloor N/2 \rfloor)} \}  }{\mathrm{argmax}} \lambda \Bigg( \frac{||q_1 - \bar{q}||_2}{\max_i ||q_{(i)} - \bar{q}||_2} + \frac{||q_2 - \bar{q}||_2}{\max_i ||q_{(i)} - \bar{q}||_2} \Bigg) 
\]
\[
\ \ \ \ \ \ \ \ \ \ \ \ \ \ \ \ \ \ \ \ \ \ \ \ \ \ \ \ \ \ \ \ \ \ \ \ \ \ \ \ \ \ \ \ \ \ \ \   -\ (1-\lambda) \Bigg( \langle \frac{q_1 - \bar{q}}{||q_1 - \bar{q}||_2} , \frac{q_2 - \bar{q}}{||q_2 - \bar{q}||_2}  \rangle + 1 \Bigg) \ .
\]
\normalfont
The first term guarantees solutions with maximum distance from median, and second term guarantees that the vector connecting the median to these points are in opposite direction (with best value -1). The interquartile distance (IQR) can be obtained by $ IQR = || \tilde{q}_{Q_1} - \bar{q} ||_{_2} + || \tilde{q}_{Q_3} - \bar{q} ||_{_2} $, and the standard cut-offs can be calculated as:
\[
\tilde{q}_{w_1} = \tilde{q}_{Q_1} + 1.5 IQR \frac{\tilde{q}_{Q_1} - \bar{q}}{|| \tilde{q}_{Q_1} - \bar{q} ||_{_2}}\ \ \ \  \text{and}\ \ \ \  \tilde{q}_{w_3} = \tilde{q}_{Q_3} + 1.5 IQR \frac{\tilde{q}_{Q_3} - \bar{q}}{|| \tilde{q}_{Q_3} - \bar{q} ||_{_2}} \ .
\]
The points in the sample set that are closest to these standard cut-offs are the extremes (provided that they are neither included in central region set nor flagged as outliers). Note them as $\tilde{q}_{e_1}$ and $\tilde{q}_{e_3}$.  A function $\tilde{f}$ is identified as outlier if $ || \tilde{q} - \bar{q} ||_{_2} > \max \{ || \tilde{q}_{e_1} - \bar{q} ||_{_2} , || \tilde{q}_{e_3} - \bar{q} ||_{_2} \} $. If we want to be less conservative,``$\max$" in the inquality can be replaced with ``$\min$". This completes the development of amplitude boxplot.

\subsection{Phase Boxplot}

Unlike the amplitude variability case for which the transformation $Q(f)$ suggested a space with linear geometry, the representation space of phase variability $\Gamma$ has a nonlinear geometry. This makes the development of phase boxplot an intricate task, as the calculations cannot be done using linear algebra. Fortunately, there is a helpful mapping called square root transformation (SRT) defined as $\psi = \sqrt{\dot{\gamma}}$, which transforms the elements of phase space $\Gamma$ to a well-known space for which there are explicit/analytical formulas that can be used for the development of boxplots \citep{srikla2016,xiekubhas2017}. Any $\gamma(t)$ can be recovered by $\gamma(t) = \int_{0}^t \psi^2(s) ds$. Obviously, $||\psi ||_{_2} = 1$ and $\psi(t) > 0$, which imply that the elements of SRT lie on the posistive orthant of the unit Hillbert sphere space $\Psi$. By using the geometry of space $\Psi$, the distance between  $\gamma_1$ and $\gamma_2$ can be obtained as $ D_p(\gamma_1,\gamma_2) = cos^{-1} (\langle \psi_1,\psi_2 \rangle)$. For the calculation of median $\bar{\psi}$, the optimization problem given in Eq.(\ref{D_a}) is used by replacing $D_a$ with $D_p$.

For the development of boxplot, the tangent space should be determined.  Tangent space at point $\psi \in \Psi$ is defined as $T_{\psi}(\Psi)=\{\nu | \langle \nu,\psi \rangle = 0,\ \nu: [0,1] \to \mathbb{R} \}$. To form the boxplot, it is needed to map all the warping functions $\{ \gamma_1,\gamma_2,...,\gamma_N \}$ to the tangent space at $\bar{\psi}$, which gives $\{ \nu_1,\nu_2,...,\nu_N \}$. This can be done using inverse exponential map:
\[
\nu_i = \exp_{\bar{\psi}}^{-1} (\psi_i) = \frac{\theta}{sin(\theta)} (\psi_i - cos(\theta) \bar{\psi}), \ \ \ \ \ \ \ \theta = cos^{-1}(\langle \psi_i,\bar{\psi} \rangle) \ , \ \ \ i=1,...,N \ .
\]
Since the tangent space is a linear space, the phase boxplot can be constructed in this space by following the steps given for amplitude boxplot. Those steps are not repeated here for the sake of brevity. After forming the boxplot in the tangent space, the last step is to transfer back the boxplot to the original space $\Psi$ (and $\Gamma$). This is done using the exponential map:
\[
\psi_i = \exp_{\bar{\psi}} (\nu_i) = cos(||\nu_i||_{_2})\bar{\psi} + sin(||\nu_i||_{_2}) \frac{\nu_i}{||\nu_i||_{_2}} \ , \ \ \ i=1,...,N \ ,
\]
 which finishes the construction of phase boxplot.

\section{Considered 3D Geometries} \label{3DModel}

In this section, the details of the 3D geometries used in the current investigation for ascertaining the veracity of the considered technique are given. The considered geometries include a benchmark geometrical shape modelled in the form of multivariate functionals, and four different 3D products extracted from a well-known database for 3D printing, called Thingi10K. The details of these 3D geometries are given in the following sub-sections.

\subsection{Benchmark Geometry}

\begin{figure}
	\begin{center}
 		\includegraphics[width=0.8\textwidth,angle =0, scale=1.2]{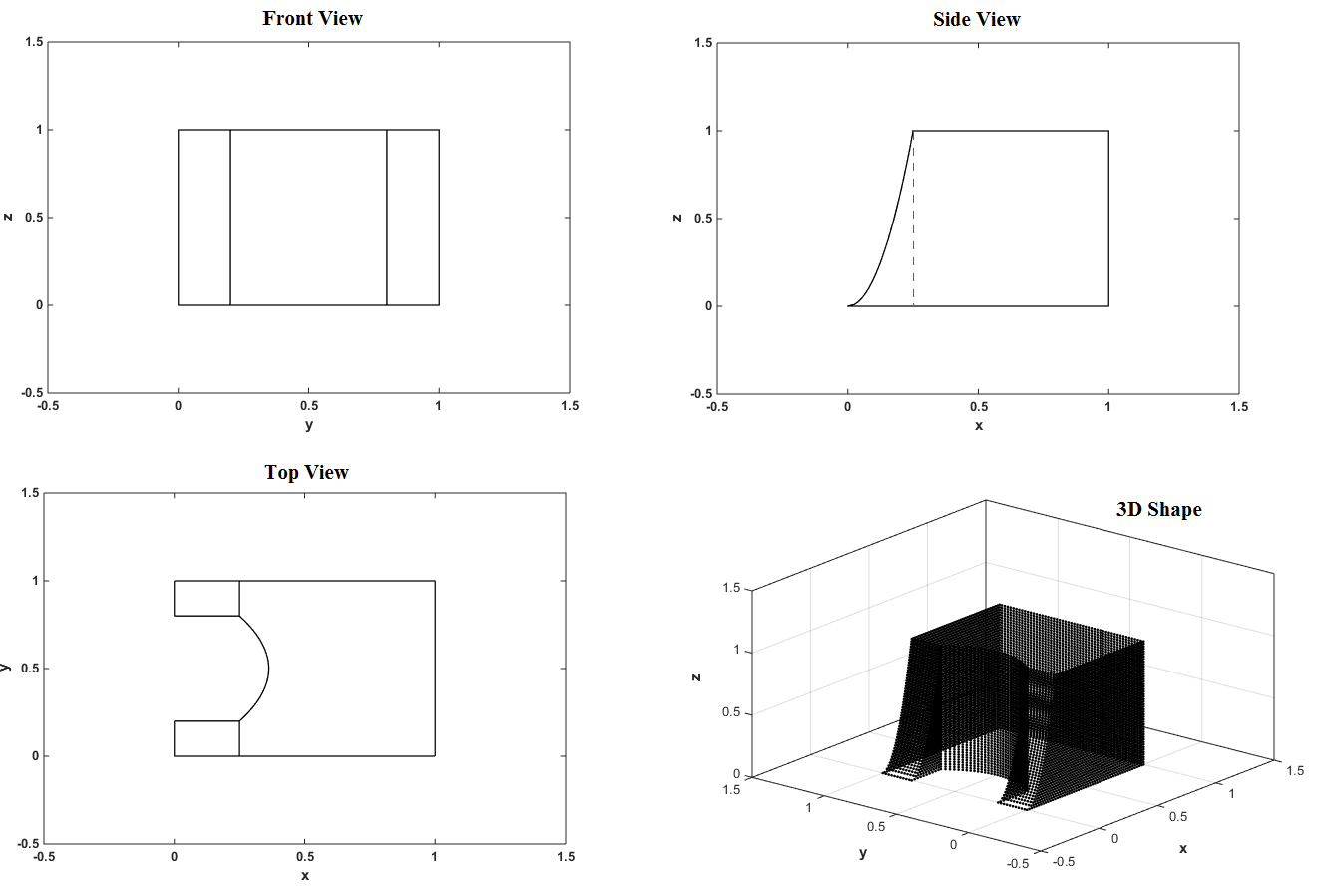}
 	\end{center} \caption{Schematic illustration of the considered geometrical shape, (a) front view, (b) side view, (c) top view, and (d) 3D shape. The values of $x$, $y$, and $z$ are given in $mm$.}\label{ShapeView}
 \end{figure}

The main reason behind considering the current benchmark geometry is that it has an exact analytical model which enables us to deliberately produce deviated shapes for investigating the capability of the considered method for detecting different types of deformation scenarios. Figure \ref{ShapeView} presents different views of the considered shape. As seen, the geometry of the shape has a non-convex boundary, and includes both linear and nonlinear curves at its sides. The multivariate functional formulation of the boundary of the 3D geometrical model is given in Appendix \ref{AA}. 

\begin{figure}
	\begin{center}
		\includegraphics[width=0.7\textwidth,angle =0, scale=1.2]{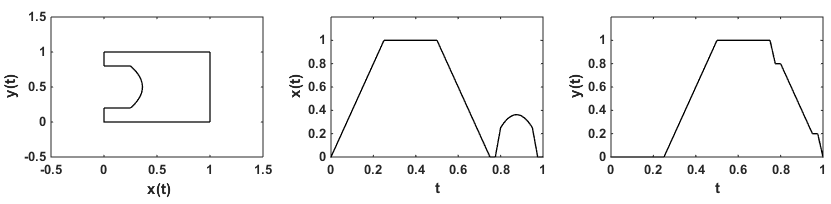}
	\end{center} \caption{Collected data for the first layer (from bottom) including in-plane contour and functionals $x(t)$ and $y(t)$ for $t \in [0,1]$, where $t$ is the shape parameter indicating the trajectory of fabrication. The values of $x$ and $y$ are given in $mm$. }\label{ContourAndFunctionals}
\end{figure}

Since 3D printing is a digital manufacturing process, the production of the above geometry is done layer-by-layer. Therefore, to make the considered statistical strategy compatible, the analysis is performed on different layers of the product, by considering the 2D contours of the original model. Figure \ref{ContourAndFunctionals} depicts the collected data for the first layer (from bottom) of the 3D product. Note that since the layer is specified, the subscript $z$ is removed and the two functionals are presented by $x(t)$ and $y(t)$. 

\subsection{Geometries from Thingi10K}

\begin{figure}
	\begin{center}
		\includegraphics[width=0.7\textwidth,angle =0, scale=1.2]{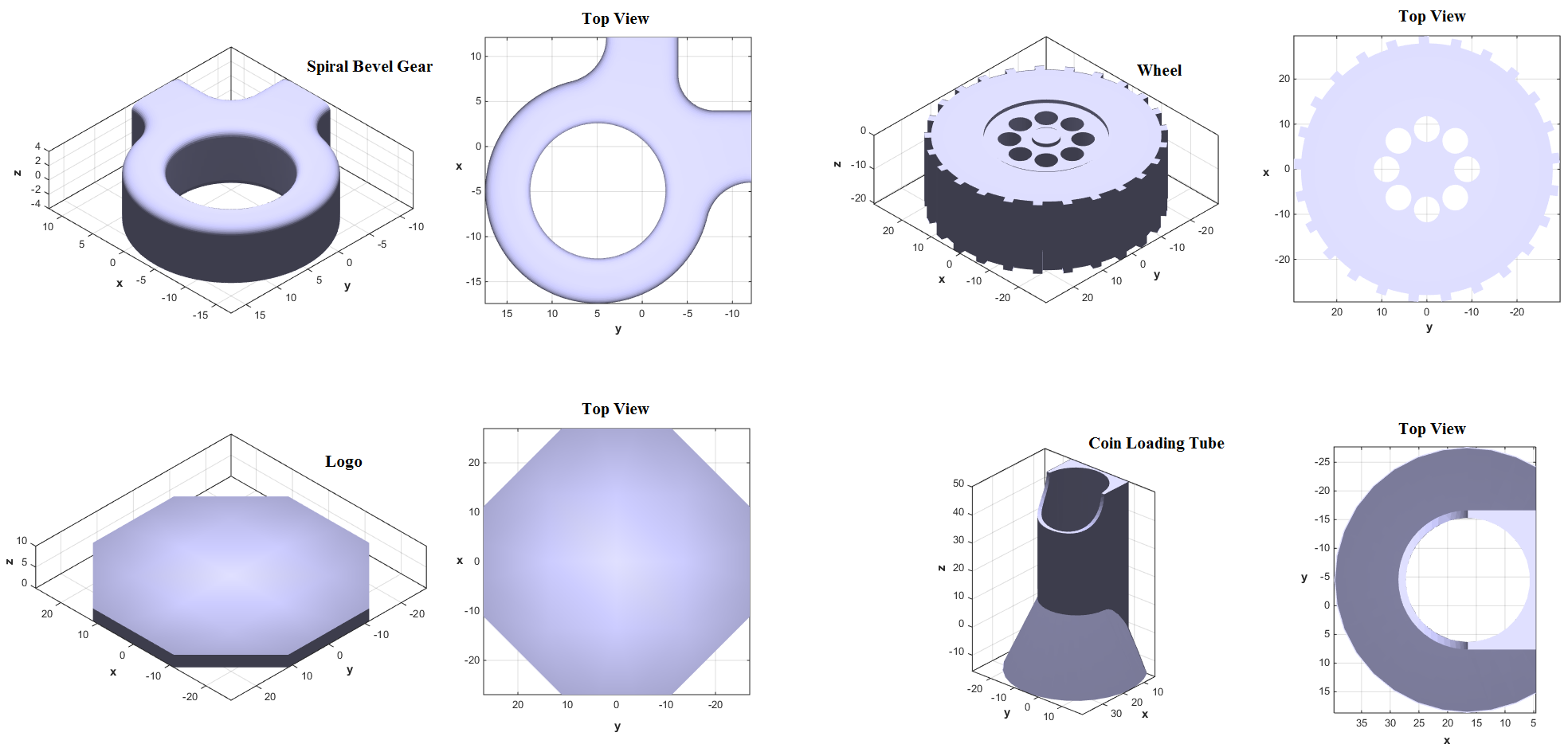}
	\end{center} \caption{Geometries obtained from Thingi10K Database. The values of $x$, $y$, and $z$ are given in $mm$.}\label{Thingi10KModels}
\end{figure}

Thingi10K is one of the largest open access databases of 3D printed models which includes 10000 datasets offering a wide range of geometrical properties \citep{zhja2016}. This versatily makes the available geometries best suited for evaluating the potential of process control algorithms. The extracted geometries are shown in Figure \ref{Thingi10KModels}, which are a spiral bevel gear (model 34785), a wheel (model 37841), a logo (model 97494),  and a coin loading tube (model 36090). As seen, the collected geometries have different shapes that are rather complicated, and can make the process monitoring and outlier detection tasks challenging. 
It is worth noting that all of the abovementioned models are downloadable in \textit{.stl} file format, and to make them prepared for layer-by-layer analysis and decomposable in the form of separate functionals $x_z(t)$ and $y_z(t)$, some preprocessing including removing tied data as well as sorting the vertices sampling flow is reqired. 

\begin{figure}
	\begin{center}
		\includegraphics[width=0.7\textwidth,angle =0, scale=1.2]{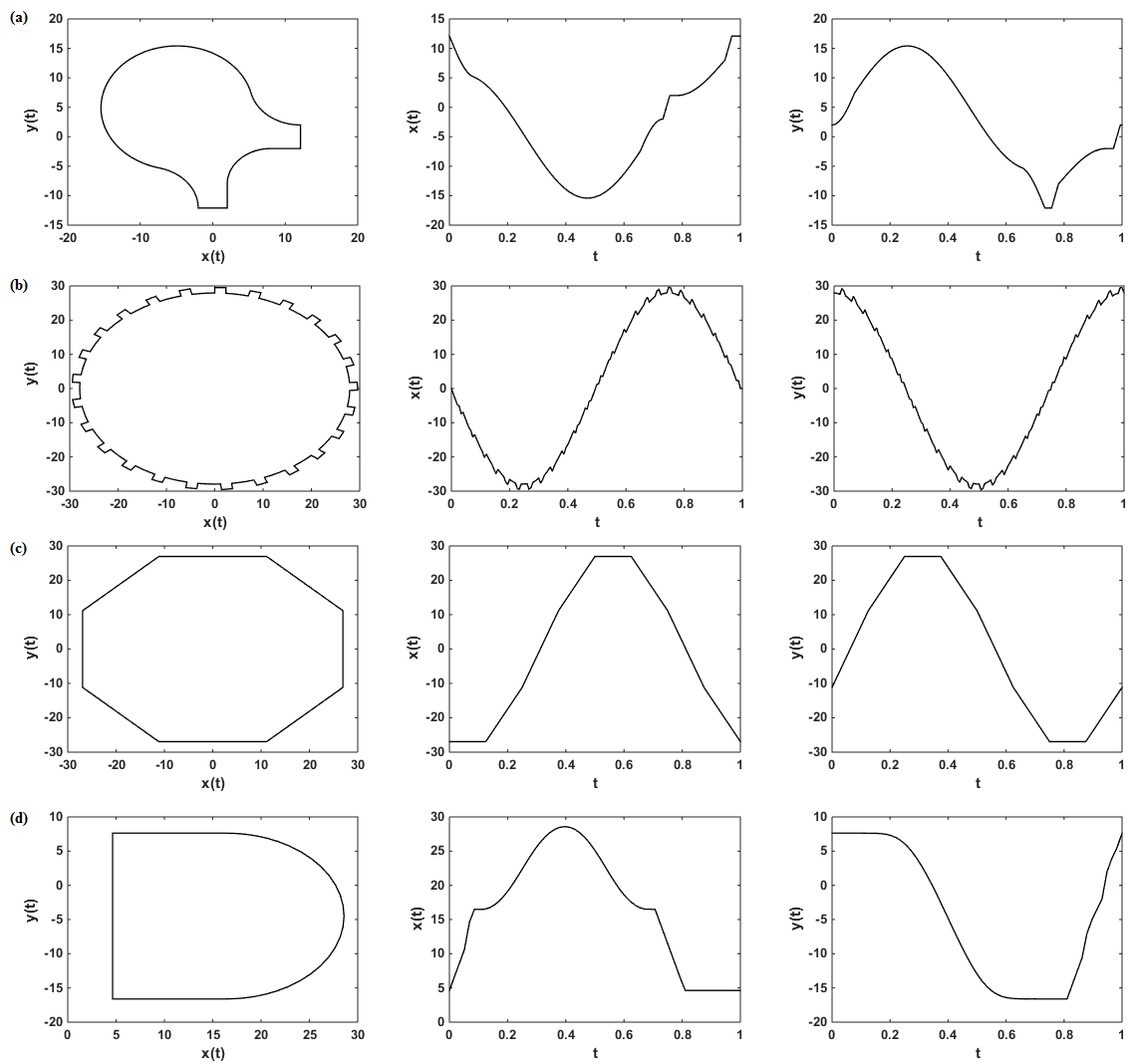}
	\end{center} \caption{External contour and corresponding functionals $x(t)$ and $y(t)$ for (a) spiral bevel gear, (b)  wheel, (c)  logo, and (d) coin loading tube. The values of $x$ and $y$ are given in $mm$.}\label{Thingi10KLayerFunctionals}
\end{figure}

The external contour of the selected layer for each of the products as well as the corresponding functionals $x(t)$ and $y(t)$ at a specified layer are shown in Figure \ref{Thingi10KLayerFunctionals}. The benefit of modeling the geometries using functionals $x(t)$ and $y(t)$ with shape parameter $t \in [0,1]$ is that it gives us enough flexibility to produce geometries with outlying phase or amplitude component to investigate the performance of the considered method. Also, if there is a need to have a uniformly sampled data (which can be the case as it is possible that real datasets are irregularly sampled) or a continuous model, Fourier series can be used for approximation.
  
\begin{figure}
	\begin{center}
		\includegraphics[width=0.7\textwidth,angle =0, scale=1.2]{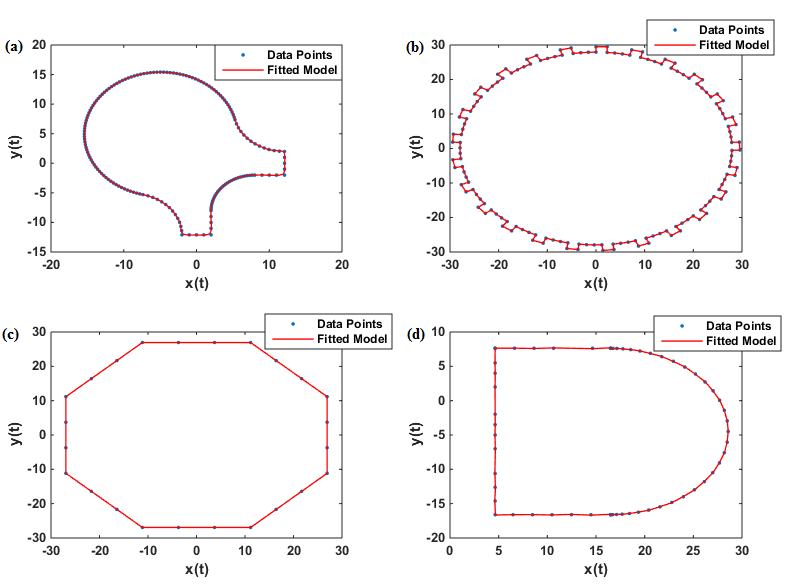}
	\end{center} \caption{Comparison of approximated contours and real data for (a) spiral bevel gear, (b)  wheel, (c)  logo, and (d) coin loading tube. The values of $x$ and $y$ are given in $mm$.}\label{Thingi10KFourierApprox}
\end{figure}

For the current simulation, it is needed to have a continuous model as a function of shape parameter $t$ to enable the generation of variability with respect to phase components using warping functions. The functional contours of the model can be approximated using the Fourier expansion. To ensure the Fourier expansion is a reliable approximation model for the contour data, a numerical study is performed. To do so, the digital information including the position of points on the external boundary of the products at the specified layer (see  Figure \ref{Thingi10KLayerFunctionals}) is collected, and then, Fourier model is fitted to the data to come up with approximations. Due to the independency of the two functionals $x(t)$ and $y(t)$, two Fourier series can be used to fit the data for producing continuous functional models from discrete data. Figure \ref{Thingi10KFourierApprox} depicts the approximated curves, where each red solid curve shows the fitted model and blue dots are points sampled from the boundary of products. The number of basis used for the approximation of spiral bevel gear, wheel, logo, and coin loading tube are 81, 149, 21, and 51, respectively. Our experiment indicates that the model has a very good performance and can be used in practice whenever no functional model is available.

\section{Monte Carlo Simulation Setup} \label{MCMCsetup}

The base geometries defined in the last section are exposed to different random deformation (obtained by MC simulation) to collect the required data for functional outlier detection. For all of the numerical experiments, 150 samples are drawn to do the analysis ($N = 150$). The simulation setup consists of three deformation scenarios which aim at testing the considered method when there are products with outlying surface roughness, amplitude component, phase component, or any other deformation resulting from possibly unseen process parameters. Each of the scenarios presented are applied to benchmark geometry as well as the four geometries obtained from Thingi10K database. The considered MC simulation performed for deforming the 3D shapes are presented below: 

\textbf{Simulation 1:} One of the most probable defective patterns of 3D products is the surface roughness which can appear as build-material uncertainty due to unstable droplet positioning, and varying droplet size and shape. Such flaws lead to geometrical defects as well as height error resembling the addition of random field noise to the surface of 3D products \citep{coli2010}. Therefore, MC simulation is conducted by adding Gaussian error to the contour curves. For benchmark geometry, from the generated samples, the first 148 are drawn under safe deformation condition, and the remaining 2 samples are drawn from an outlying distribution. The mathematical formulation of noises for safe and significant deformations are given in Appendix \ref{AA}. For the four Thingi10K geometries, a similar simulation setup is followed. The mathematical formulation as well as the specified values of the safe and outlying distributions of roughness noises are presented in Appendix \ref{AA}. For all shapes, it is expected that 1 or 2 outliers with significant surface roughness be detected.

\textbf{Simulation 2:} The other common reason behind geometrical deformation in 3D printing is in-situ environmental uncertainty. In this condition, the uncontrolled fluctuations of build-envelope temperature incur melting and results in the surface deformation. Inline with this, the purpose of the this simulation is to evaluate the performance of the considered statistical method in detecting significant outliers with respect to amplitude component. For the benchmark geometry, the robustness of the considered statistical method is tested for detecting outliers when there is an insignificant random time warping between the deformed shapes. To this aim, a waveform (periodic) deformation is considered in which some of the data may have significant deformation with respect to the amplitude. Also, there is a random time warping in the deformed curves with no phase outlier. It is expected that the considered outlier detection method withstand against insignificant random time warping and identify the outliers due to significant amplitude deformation, thanks to the decomposition of functional data. For all of the samples, the shape deformation is imposed on side $t \in t_1$. The mathematical formulation of the deformation is given in Appendix \ref{AA}. For the four Thingi10K geometries, simulation study is carried out to evaluate the performance of the method for detecting 3D products with outlying amplitude component.  The mathematical formulation of the simulation scenario can be found in Appendix \ref{AA}.  For all shapes, it is expected that 3 to 4 products with significant outlying amplitudes be detected.

\textbf{Simulation 3:} Malfunctions such as improper performance of powder deposition system, scanning strategy, and incorrect air gas flow in AM can result in significant warping defects \citep{cohudats2018}. Therefore, it is necessary to make sure such outliers can be detected. This corresponds to the detection of 3D products with outlying phase component. For the benchmark geometry, a uniform periodic deformation is inflicted on side $t \in t_1$, as given in Appendix \ref{AA}. The mathematical details for the four Thingi10K geometries are also presented in  Appendix \ref{AA}.  For all shapes, it is expected that 3 to 4 samples be significantly different from the others with respect to phase component.

\begin{figure}
	\begin{center}
		\includegraphics[width=0.8\textwidth,angle =0, scale=1.1]{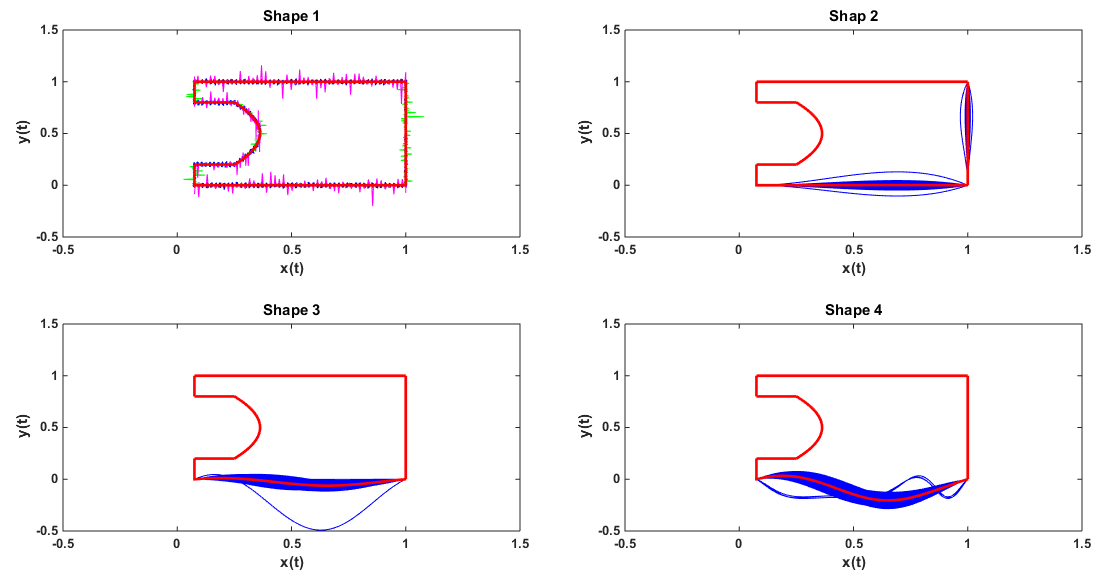}
	\end{center} \caption{The collected contour data of $10^{th}$ layer of geometrical shapes for different simulations of benchmark geometries, red profiles are sample means and the other profiles are the samples. The values of $x$ and $y$ are given in $mm$.}\label{ContourShapesData}
\end{figure}

\begin{figure}
	\begin{center}
		\includegraphics[width=0.8\textwidth,angle =0, scale=1.1]{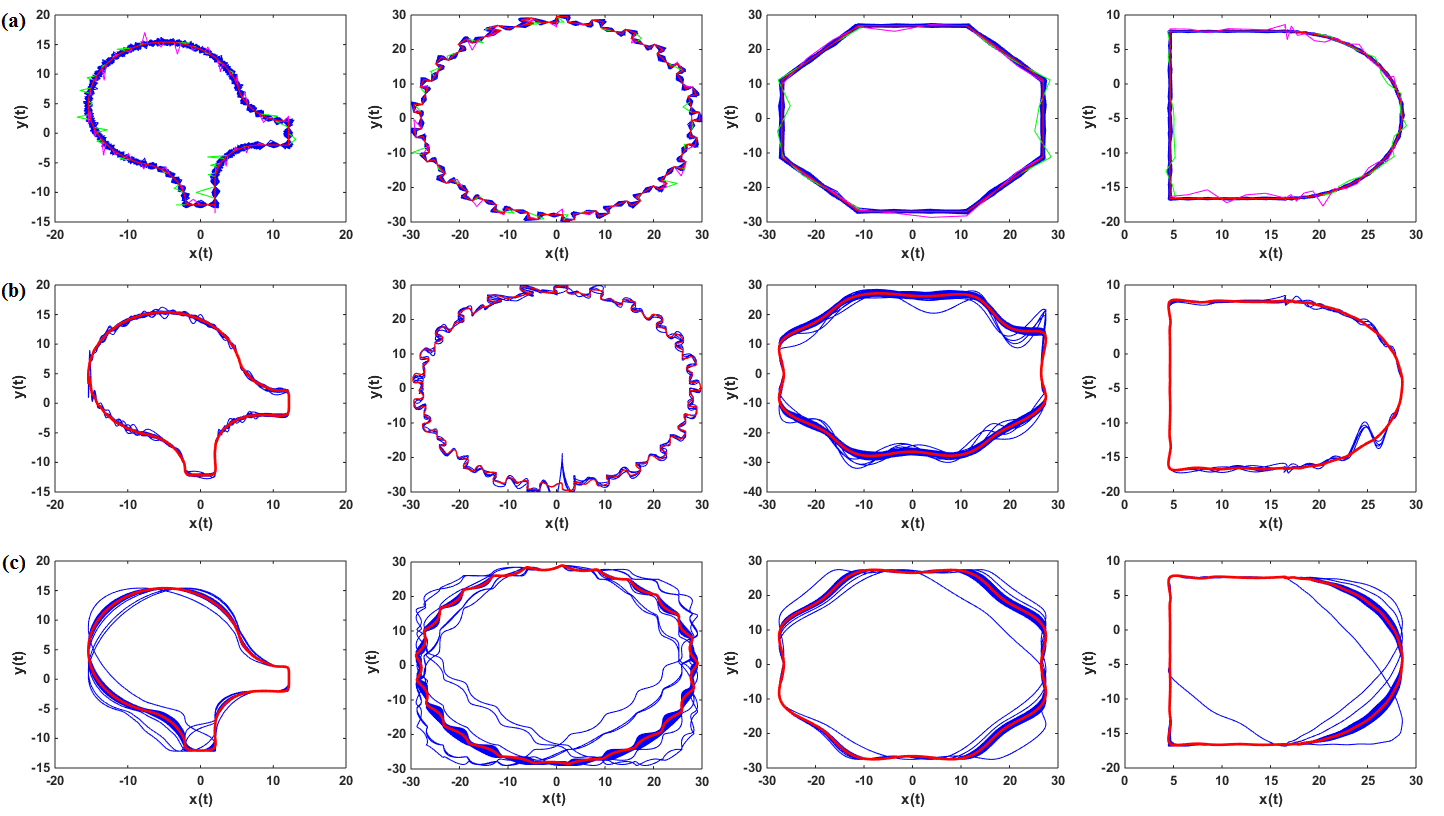}
	\end{center} \caption{Sample shapes for different simulations of Thingi10K geometries, red profiles are sample means and the other profiles are the samples. The values of $x$ and $y$ are given in $mm$.}\label{Thingi10KSimulationData}
\end{figure}

Figure \ref{ContourShapesData} depicts the contour data together with the sample contour mean for each of the simulation case studies of benchmark geometries. Figure \ref{Thingi10KSimulationData} shows the contour data as well as the sample contour mean for each of the simulation cases of Thingi10K geometries. In these figures, each row corresponds to a simulation scenario. The sample mean is shown by red and the collection of sample contours are shown by blue. It can be seen that in the collected samples, some shapes have statistically significant deviation from the sample mean geometries and there is enough variablity in the samples which enables us study the performance of the considered outlier detection method. It is also worth mentioning that we assume these samples are collected from different AM production runs to include enough variablity in the collected samples, as the main objective of the considered study is to evaluate the performance of the method for outlier detection.

\section{Simulation Results} \label{simresults}

\subsection{Benchmark Geometry Case Studies} 
\begin{figure}[!h] 
	\begin{center}
		\includegraphics[width=0.75\textwidth,angle =0, scale=1]{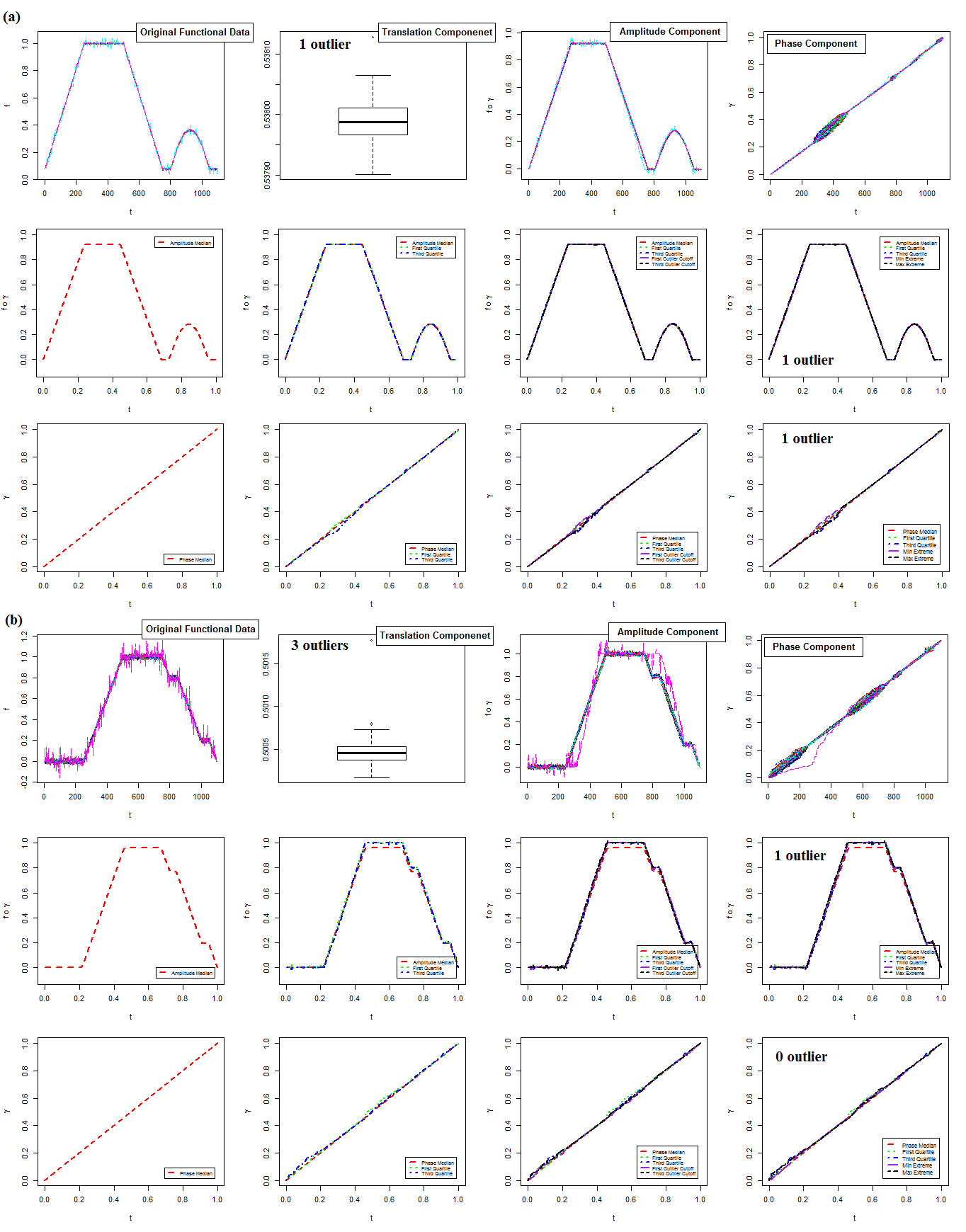}
	\end{center} \caption{Results of simulation 1 of benchmark geometries for (a) $x(t)$ and (b) $y(t)$.}\label{Shape1}
\end{figure}

This sub-section is devoted to the discussion of the obtained results for simulations 1 to 3 of benchmark geometries. The functional boxplots as well as the decomposed components of $x(t)$ and $y(t)$ for simulation 1 of benchmark geometries are depicted in Figure \ref{Shape1}. For the functional boxplots of benchmark geometries, the results for each independent analysis (e.g. $x(t)$ in part a and $y(t)$ in part b of  Figure \ref{Shape1}) are given in three rows. The first row depicts the original functional samples together with translation boxplot, amplitude components and phase components obtained from decomposition of phase and amplitude components. The second row indicates the amplitude median, first and third quartiles, first and third outlier cutoffs and the resulting amplitude boxplot. The third row contains the same information for the phase component. It can be realized from the boxplots that the outlying contours are correctly identified. One interesting feature of this simulation is that roughness noises do not necessarily follow a periodic or well-behaved pattern, which can make the detection challenging. However, the decomposition strategy followed yields promising results and detects the two products with significant surface roughness. This suggests the acceptable performance of the considered approach even when the functionals being analyzed are non-smooth, although the method is originally designed to work with smooth functionals (see definition of $q$ in sub-ection \ref {setnotations}). 

Figure \ref{Shape3} depicts the results of simulation 2 of benchmark geometries. As seen, the obtained boxplots have a nicer form which is because this simulation is particularly designed for evaluating the accuracy of the considered method for detecting outliers in the amplitude component. For this simulation scenario, 1 outlier is detected from translation boxplot, and 2 outliers are detected from amplitude boxplot (one of them is in common with translation boxplot), while no outlier is detected from phase boxplot. This is completely inline with the setup of simulation 2 in which the goal was to identify outliers in the presence of uniform random time warping. The observed results ensure the robustness of the considered outlier detection method when there is additional insignificant phase variability. 
	
The results obtained for simulation 3 of benchmark geometries are given in Figure \ref{Shape4}. Again, the obtained boxplots and the decomposition and aligning of the original functionals is clear, as the study is designed for testing the performance of the method when there are outlying phase components. For this case, 4 outliers are detected from which 3 outliers are commonly identified from both phase and amplitude boxplots, and one additional outlier belongs to phase boxplot. Also, there is no translation outlier in the data. From the aligned functions as well as the way the amplitude component of functionals are produced, it can be realized that there is a chance to get some functionals with outlying amplitude component, even if the main purpose was to check for phase outliers. Moreover, it can be inferred that the obtained results are reliable, as all of the outlying products are also detected by phase boxplot. This attests the validity of the method in detecting products with outlying phase components. 

The findings of these experiments show that the considered method is useful for data cleaning task which is a standard stage followed by quality control engineers to come up with a reliable control chart through phase I analysis. 

\begin{figure}[!h] 
	\begin{center}
		\includegraphics[width=0.8\textwidth,angle =0, scale=1.2]{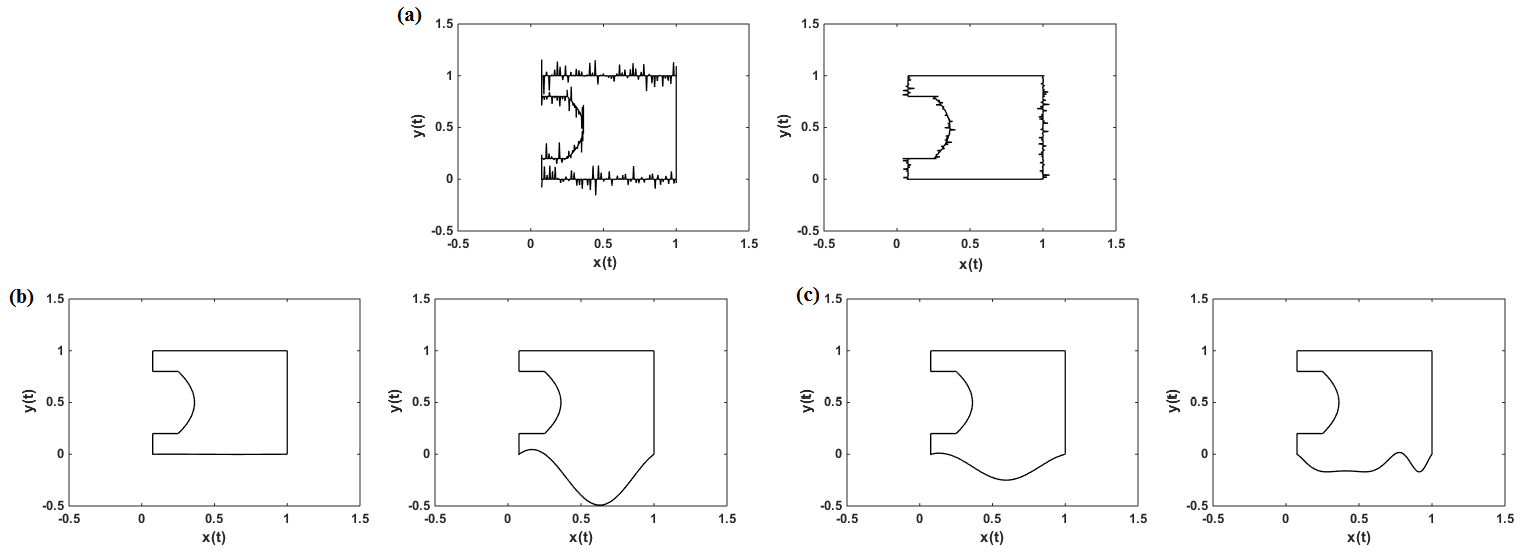}
	\end{center} \caption{Some of the detected contours for (a) simulation 1, (b) simulation 2, and (c) simulation 3. The values of $x$ and $y$ are given in $mm$.}\label{OutlierSim1234}
\end{figure}

To give the readers a clear sense on how successful the methods were in detecting significantly deformed products, the contours of some of these products at the selected layer are presented in Figure \ref{OutlierSim1234}. It is clear from the geometries that the method has high competence to be used for process monitoring in AM applications, as the identified products have obvious defective geometrical features, and none of such detections can be regarded as a false alarm.

\subsection{Thingi10K Case Studies} At the second stage of the numerical experiments, the considered method is applied to a more challenging outlier detection task for further evaluations. As can be inferred from Figure \ref{Thingi10KLayerFunctionals}, the selected geometries have different types of features. For example, the contour of wheel model is quite non-smooth which allows us to test the robustness of the considered method when it is used for monitoring products with complicated shapes. Also, the authors believe that the geometrical features offered by Thingi10K geometries are enough for producing any arbitrary geometry, and thus, the current simulation provides us with some judgement regarding the applicability of the method for monitoring of products with freeform geometries.

 Figure \ref{Sim5PlotsThingi10K} shows the results of simulation 1 for Thingi10K geometries (for the sake of brevity, all of the functional boxplots obtained from Thingi10K geometries analysis only provide the boxplots for translation, amplitude and phase components). The boxplots detect 4 outliers for spiral bevel gear case, 3 outliers for wheel case, 4 outliers for logo case, and 5 outliers for coin loading tube case. For all of the cases, the samples drawn from significant distributions are correctly detected. Also, it happens that some of the products generated from safe distribution are categoried as outlier, which is reasonable as there is always a possibility that Gaussian distribution generates outlying samples. For further elaboration, some of the detected geometries for each of the four models are depicted in Figure \ref{OutlierSim5}. It is apparent that the detected shapes have significant deformation from the original model, and functional boxplots yield promising results.

Figure \ref{Sim6PlotsThingi10K} presents the results of simulation 2 for Thingi10K geometries.  7 outliers are detected for spiral bevel gear case, 4 outliers are detected for wheel case, 4 outliers are detected for logo case, and 5 outliers are detected for coin loading tube case. The results of spiral bevel gear and wheel cases are very promising, as the outliers are all detected by the amplitude boxplot. For the other two cases, the amplitude boxplot detects all of the deformed shapes, but those shapes are also detected by phase boxplot. The obtained results also show that the translation boxplot properly detects the outlying products. Since the amplitude boxplot is successfull for all of these cases, the overall observation is that functional boxplots can properly handle the detection of outliers in the amplitude component of the products. It should be noted that for two of the cases, the phase boxplots also send the false alarm, which may not be so desirable, but still does not degenerate the performance of the method, as none of these detected products have acceptable quality (they are outliers with respect to amplitude as well). Figure \ref{OutlierSim6} depicts some of the detected shapes for this simulation, which further endorses the applicability of the method. 

Figure \ref{Sim7PlotsThingi10K} depicts the functional boxplots for simulation 3 for Thingi10K geometries. The results for this simulation scenario is very promising, as for all of the cases the outlying products are detected by translation and phase boxplots (no outlier is detected by amplitude boxplot). The boxplots yield 3 outliers for bevel gear case, 7 outliers for wheel case, 2 outliers for logo case, and 3 outliers for coin loading tube case. It seems that for three of the geometries, the phase boxplot cannot detect all of the outlying products (1 undetected product for logo case, and 2 undetected products for wheel and coin loading tube cases), yet those undetected products are identified by translation boxplots. Figure \ref{OutlierSim7} shows some of the detected products. It can be seen that the detected products have obvious deformations from corresponding standard models. 

All in all, it can be inferred that the considered method does a good job in the detection of outlying products for the four Thingi10K models, and the robustness of the method is not affected by the geometrical complexity associated with their shapes.

\section{Discussions} \label{discussions}

In this section, based on the authors' evaluation supported by numerical experiments, some general observations regarding the properties and potentials of the considered method are given. It is believed that these points can be useful for applying the adopted functional boxplot for other AM applications. 

\begin{itemize}
	
	\item From the simulation results (in particular simulation 2), it was observed that the implemented method is capable of data cleaning which is an important step taken by quality control engineers during phase I analysis. This feature together with the acceptable performance of the method in detecting deformed products give it merit for practical quality assurance and process monitoring applications.

	\item An interesting property of the considered functional boxplot is the visualization of function alignment and corresponding decompositions. This graphical toolbox gives the user a clear view regarding the nature of the outlier detection process for a complex task like outlier detection in 3D products. Also, this toolbox can be viewed as an approach that automatically yields interpretable results, understandable even by practitioners with elementary statistical background. This feature is advantageous in a sense that it makes this boxplot an easy to use tool for practitioners, despite the fact that lots of formidable technical and theoretical concepts from the fields of topology, smooth manifolds, statistics and functional analysis are used for its prepration.

\item As pointed out recently in \cite{cohudats2018}, the geometrical defects in AM can be resulted from different parameters such as  improper performance of powder deposition system, scanning strategy, incorrect air gas flow, etc. Such malfunctions not only can affect the amplitude component of the geometries, but also can result in obvious warping defects. Therefore, the authors believe that the considered toolbox is best suited for monitoring such defects, and can be considered as a solution for the future of process monitoring in AM industry.

\item In almost all of the considered simulation cases, it was observed that the deformed shapes were detected correctly, and there were no healthy products among them. So, it is expected that the considered method has a good performance with respect to false alarm rate measure, and can be safely used by AM manufacturers. 

\item AM process monitoring is among those tasks which demand computationally efficient techniques. The layer-by-layer analysis proposed in the current research turns the process monitoring of 3D geometries into a 2D contour analysis. In all of the considered simulation cases, the authors were successful to use two separate Fourier series models for approximating $x(t)$ and $y(t)$ functionals which reduced the complex problem of monitoring 3D geometries to profile monitoring problem. It is believed that, whenever possible, this strategy should be followed for the analysis. Otherwise, it may be needed to use a large data cloud to present the whole geometry, for which the computation requirement of the analysis becomes truely intensive.

\end{itemize}	

\color{black}

\section{Conclusion and Future Work} \label{conclusions}

In the current study, a novel statistical technique was proposed for the offline/online detection of deformed 3D products during the additive manufacturing process. Through different MC simulation, it was observed that the considered statistical outlier detection method can successfully discriminate 3D products with statistically significant deformation. In particular, decomposing the contour functional data into translation, amplitude, and phase components and developing functional boxplots for each of them was found to be very beneficial in terms of robustness. In future, a multivariate functional outlier detection method will be developed to study much more intricate geometrical shapes for which it is not safe to find separate functional approximations for $x(t)$ and $y(t)$. This allows extending the area of application of the considered method from layer-by-layer contour analysis to surface analysis of possibly complicated geometries. Also, a theoretical analysis will be conducted to investigate the assymptotic properties of the considered outlier detection technique.

\bibliographystyle{apalike}
\bibliography{References.bib}

\newpage
\appendix

\section{Mathematical Formulation of Benchmark Geometry and Simulation Scenarios} \label{AA}

\subsection{Benchmark Geometry Model} The mathematical formulation of the benchmark geometry is given below:

\[
x_z(t) = \begin{cases} \tilde{x}_1(z) + (\cfrac{\tilde{x}_2(z) - \tilde{x}_1(z)}{\max(t_1)-\min(t_1)})t & \mbox{if } t \in t_1 \\ 
\tilde{x}_2(z) & \mbox{if } t \in t_2 \\  
\tilde{x}_3(z) - (\cfrac{\tilde{x}_3(z) - \tilde{x}_4(z)}{\max(t_3)-\min(t_3)})(t-\min(t_3)) & \mbox{if } t \in t_3 \\ 
\tilde{x}_4(z) & \mbox{if } t \in t_4 \\ 
\tilde{x}_5(z) + (\cfrac{\tilde{x}_6(z) - \tilde{x}_5(z)}{\max(t_5)-\min(t_5)})(t-\min(t_5)) & \mbox{if } t \in t_5 \\  
\tilde{x}_6(z)+0.1125-20(t-\cfrac{\max(t_6)-\min(t_6)}{2})^2 & \mbox{if } t \in t_6 \\ 
\tilde{x}_7(z) - (\cfrac{\tilde{x}_7(z) - \tilde{x}_8(z)}{\max(t_7)-\min(t_7)})(t-\min(t_7)) & \mbox{if } t \in t_7 \\ 
\tilde{x}_8(z) & \mbox{if } t \in t_8 \end{cases}  \   ,
\]
\[
y_z(t) = \begin{cases} \tilde{y}_1(z) & \mbox{\ \ \ \ \ if } t \in t_1 \\ 
\tilde{y}_2(z) + (\cfrac{\tilde{y}_3(z) - \tilde{y}_2(z)}{\max(t_2)-\min(t_2)})(t-\min(t_2)) & \mbox{\ \ \ \ \ if } t \in t_2 \\  
\tilde{y}_3(z) & \mbox{\ \ \ \ \  if } t \in t_3 \\ 
\tilde{y}_4(z)-(\cfrac{\tilde{y}_4(z)-\tilde{y}_5(z)}{\max(t_4)-\min(t_4)})(t-\min(t_4)) & \mbox{\ \ \ \ \ if } t \in t_4 \\
\tilde{y}_5(z) & \mbox{\ \ \ \ \ if } t \in t_5 \\  
\tilde{y}_6(z)-(\cfrac{\tilde{y}_6(z)-\tilde{y}_7(z)}{\max(t_6)-\min(t_6)})(t-\min(t_6)) & \mbox{\ \ \ \ \ if } t \in t_6 \\ 
\tilde{y}_7(z) & \mbox{\ \ \ \ \ if } t \in t_7 \\ 
\tilde{y}_8(z)-(\cfrac{\tilde{y}_8(z)-\tilde{y}_1(z)}{\max(t_8)-\min(t_8)})(t-\min(t_8)) & \mbox{\ \ \ \ \ if } t \in t_8 \end{cases}  \   , 
\]
\[
z(s)=s; \ \ s \in [0,1] \ , \ \ \ \ \ \ \ \ \ \ \ \ \ \ \ \ \ \ \ \ \ \ \ \ \ \ \ \ \ \ \ \ \ \ \ \ \ \ \ \ \ \ \ \ \ \ \ \ \ \ \ \ \ \ \ \ \ \ \ \ \ \ 
\]
where $t_1 \in [0,0.25]$, $t_2 \in [0.25,0.5]$, $t_3 \in [0.5,0.75]$, $t_4 \in [0.75,0.775]$, $t_5 \in [0.775,0.8]$, $t_6 \in [0.8,0.95]$, $t_7 \in [0.95,0.975]$, and $t_8 \in [0.975,1]$. The considered shape is a special case of the above multivariate functional model satisfying $\tilde{\mathbf{x}}(z) = (0.25 \sqrt{z},1,1,0.25 \sqrt{z}$, $0.25 \sqrt{z},0.25,0.25,0.25 \sqrt{z})^T$, and $\tilde{\mathbf{y}}(z) = (0,0,1,1,0.8,0.8,0.2,0.2)^T$.

\subsection{Simulation 1} Benchmark Geometry: The mathematical formulation of safe and significant noises are given below, respectively:
\[ \mathtt{safe:} \ 
\begin{bmatrix}
\epsilon^{(i)}_{x}(t) \\ \epsilon^{(i)}_{y}(t)
\end{bmatrix} \sim \mathcal{MVN} \Bigg( \begin{bmatrix}
0 \\ 0 
\end{bmatrix} , \begin{bmatrix}
5 \times 10^{-6} & 5 \times 10^{-6} x(t)y(t) \\ 5 \times 10^{-6} x(t)y(t) & 9 \times 10^{-5}
\end{bmatrix}\Bigg) \ , \ \ \ \ i = 1,2,...,N-2 \ ,
\]
\[ \mathtt{outliers:} \ 
\epsilon^{(N-1)}_{x}(t)  \sim \mathcal{N} \big( 0 , 0.0005 \big) \ , \ \ \ \ \ \ 
\epsilon^{(N)}_{y}(t) \sim \mathcal{N} \big( 0 , 0.0005 \big) \ ,
\]
where the last two distributions correspond to the two outliers. Once the error noises are determined, the sample contours can be obtained as $x_z^{(i)}(t)=x(t)+\epsilon^{(i)}_{x}(t)$ and $y_z^{(i)}(t)=y(t)+\epsilon^{(i)}_{y}(t)$, for $i = 1, 2, ..., N$. Also, the identity time warping function $\gamma_i(t)=t$ is considered in this simulation to further emphasize on detecting translation and amplitude outliers. \\

Thingi10K Geometries: The mathematical formulation of safe and outlying distributions of roughness noises for spiral bevel gear and  wheel are given below:
\[ \mathtt{safe:} \ 
\begin{bmatrix}
\epsilon^{(i)}_{x}(t) \\ \epsilon^{(i)}_{y}(t)
\end{bmatrix} \sim \mathcal{MVN} \Bigg( \begin{bmatrix}
0 \\ 0 
\end{bmatrix} , \begin{bmatrix}
5 \times 10^{-2} & 5 \times 10^{-5} x(t)y(t) \\ 5 \times 10^{-5} x(t)y(t) & 5 \times 10^{-2}
\end{bmatrix}\Bigg) \ , \ \ \ \ i = 1,2,...,N-2 \ ,
\]
\[ \mathtt{outliers:} \ 
\epsilon^{(N-1)}_{x}(t)  \sim \mathcal{N} \big( 0 , 2 \big) \ , \ \ \ \ \ \ 
\epsilon^{(N)}_{y}(t) \sim \mathcal{N} \big( 0 , 2 \big) \ .
\]
The safe distribution for logo is the same as above and  its outlying distribution is:
\[ \mathtt{outliers:} \ 
\epsilon^{(N-1)}_{x}(t)  \sim \mathcal{N} \big( 0 , 1 \big) \ , \ \ \ \ \ \ 
\epsilon^{(N)}_{y}(t) \sim \mathcal{N} \big( 0 , 1 \big) \ .
\]
The safe and outlying distributions for coin loading tube is:
\[ \mathtt{safe:} \ 
\begin{bmatrix}
\epsilon^{(i)}_{x}(t) \\ \epsilon^{(i)}_{y}(t)
\end{bmatrix} \sim \mathcal{MVN} \Bigg( \begin{bmatrix}
0 \\ 0 
\end{bmatrix} , \begin{bmatrix}
5 \times 10^{-3} & 5 \times 10^{-6} x(t)y(t) \\ 5 \times 10^{-6} x(t)y(t) & 5 \times 10^{-3}
\end{bmatrix}\Bigg) \ , \ \ \ \ i = 1,2,...,N-2 \ ,
\]
\[ \mathtt{outliers:} \ 
\epsilon^{(N-1)}_{x}(t)  \sim \mathcal{N} \big( 0 , 0.25 \big) \ , \ \ \ \ \ \ 
\epsilon^{(N)}_{y}(t) \sim \mathcal{N} \big( 0 , 0.25 \big) \ .
\]
The interpretation of the resulting shape deformation is the same as that of benchmark geometry.

\subsection{Simulation 2} Benchmark Geometry: The mathematical formulation of the considered deformation is given below:
\[
\begin{cases}  \begin{cases}  x_z^{(i)}(t) =x(t) \\ y_z^{(i)}(t) = \tilde{y}_1(z) + u_1 \sin(\frac{2 \pi t}{\max(t_1)}) + u_2 \cos(\frac{2 \pi t}{\max(t_1)}) - u_2 ; \ t \in t_1 \\
 y_z^{(i)}(t) =y(t); \ t \notin t_1 \end{cases}  & \mbox{if } b_i = 1 \\ \begin{cases} x_z^{(i)}(t) =x(t) \\ y_z^{(i)}(t) = \tilde{y}_1(z) + u_3 \sin(\frac{2 \pi t}{\max(t_1)}) + u_4 \cos(\frac{2 \pi t}{\max(t_1)}) - u_4 ; \ t \in t_1 \\
 y_z^{(i)}(t) =y(t); \ t \notin t_1  \end{cases}   & \mbox{if } b_i = 0 \end{cases}  \ , \ \ \ \ \ \ \ \ \ 
\]
for $i = 1, 2, ..., N$, where $b_i \sim Bernoulli(0.97)$, $\{u_1, u_2\} \sim Unif(0,0.05)$, and  $\{u_3, u_4\} \sim Unif(0.1,0.25)$. The above formulation implies that the original 3D product has naturally a periodic deformation form on side $t \in t_1$. Also, phase variability is uniformly exposed to all of the samples using the warping function $\gamma_i(t)= t + \alpha_i t(t-max(t_1))$, where $i = 1, 2, ..., N$, $t \in t_1$, and $\alpha_i \sim Uni(-1,1)$, and setting $y_z^{(i)}(t) = y_z^{(i)}(\gamma_i(t))$. \\

Thingi10K Geometries: Generating samples with random amplitude components can be done by modifying the Fourier coefficients of the fitted model. For the current simulation, the following mathematical formulation is used:
\[
\begin{cases}  \begin{cases}  x^{(i)}(t) =x(t) \\ y^{(i)}(t) = A_0 + \sum\limits_{k=1}^{K} (A_k+u_1) cos(kt) + (B_k+u_2) sin(kt) \end{cases}  & \mbox{if } b_i = 1 \\ \begin{cases}  x^{(i)}(t) =x(t) \\ y^{(i)}(t) = A_0 + \sum\limits_{k=1}^{K} (A_k+u_3) cos(kt) + (B_k+u_4) sin(kt) \end{cases}   & \mbox{if } b_i = 0 \end{cases}  \ , \ \ \ \ \ \ \ \ \ \ \ \ \ \ \ \ \ \ 
\]
for $i = 1, 2, ..., N$, where $b_i \sim Bernoulli(0.97)$, $A_0$, $A_k$ and $B_k$ are the coefficients of fitted Fourier series, and $\{u_1, u_2,u_3,u_4\}$ are uniformly distributed random variables. The number of basis $K$ for the approximation of spiral bevel gear, wheel, logo, and coin loading tube are 40, 74, 10, and 25, respectively. For spiral bevel gear case $\{u_1, u_2\} \sim Unif(0,0.05)$, and  $\{u_3, u_4\} \sim Unif(0,0.2)$,  for wheel case $\{u_1, u_2\} \sim Unif(0,0.005)$, and  $\{u_3, u_4\} \sim Unif(0,0.2)$, for logo case $\{u_1, u_2\} \sim Unif(0,0.5)$, and  $\{u_3, u_4\} \sim Unif(0,2)$, and for coin loading tube case, $\{u_1, u_2\} \sim Unif(0,0.01)$, and  $\{u_3, u_4\} \sim Unif(0,0.2)$.

\subsection{Simulation 3} Benchmark Geometry: The mathematical formulation of the considered deformation is given below:
 \[
\begin{cases}   x_z^{(i)}(t) =x(t) \\ y_z^{(i)}(t) = \tilde{y}_1(z) + u_1 \sin(\frac{2 \pi t}{\max(t_1)}) + u_2 \cos(\frac{2 \pi t}{\max(t_1)}) - u_2 ; \ t \in t_1 \\
y_z^{(i)}(t) =y(t); \ t \notin t_1   \end{cases}   \ , \ \ \ \ \ \ \ \ \ \ \ \ \ \ \ \ \ \ \ \ \ \ \ \ \ \ 
\]
for $i = 1, 2, ..., N$, where $\{u_1, u_2\} \sim Unif(0.05,0.12)$. Just like simulation 2, the above formulation means that the original 3D product has naturally a periodic deformation form on side $t \in t_1$. However, unlike simulation 2, these deformations cannot be considered as outlier. The phase variation is simulated using the following formulation:
\[
\begin{cases}  \gamma_i(t)= t + \alpha t(t-max(t_1))   & \mbox{if } b_i = 1 \\  \gamma_i(t)= t + \beta t(t-max(t_1))   & \mbox{if } b_i = 0 \end{cases}  \ , \ \ \ \ \ \ \ \ \ \ \ \ \ \ \ \ \ \ \ \ \ \ \ \ \ \ \ \ \ \ \ \ \ \ \ \ \ \ \ \ \ \ \ \ \ \ \ \ \ \ \ \ \ \ \ \ \ \ \ 
\]
for $i = 1, 2, ..., N$, and $t \in t_1$, where $b_i \sim Bernoulli(0.97)$, $\alpha \sim Unif(-0.2,0.2)$ and $\beta \sim Unif(14.3, 14.5)$, and setting $y_z^{(i)}(t) = y_z^{(i)}(\gamma_i(t))$. \\

Thingi10K Geometries: To generate samples with random phase components, the fitted Fourier models are used. The phase variation is produced by:
\[
\begin{cases}  \gamma_i(t)= t + \alpha t(t-1)   & \mbox{if } b_i = 1 \\  \gamma_i(t)= t + \beta t(t-1)   & \mbox{if } b_i = 0 \end{cases}  \ , \ \ \ \ \ \ \ \ \ \ \ \ \ \ \ \ \ \ \ \ \ \ \ \ \ \ \ \ \ \ \ \ \ \ \ \ \ \ \ \ \ \ \ \ \ \ \ \ \ \ \ \ \ \ \ \ \ \ \ \ \ \ \ \ \ \ \ \ 
\]
for $i = 1, 2, ..., N$, where $b_i \sim Bernoulli(0.97)$, and setting $y^{(i)}(t) = y^{(i)}( \gamma_i(t))$. For spiral bevel gear and logo cases $\alpha \sim Unif(-0.05,0.05)$ and $\beta \sim Unif(-0.3, 0.3)$, and for wheel and coin loading tube cases $\alpha \sim Unif(-0.05,0.05)$ and $\beta \sim Unif(-0.5, 0.5)$.

\newpage
\section{Figures from Simulation} \label{BB}

\begin{figure}[!h] 
	\begin{center}
		\includegraphics[width=0.8\textwidth,angle =0, scale=1.22]{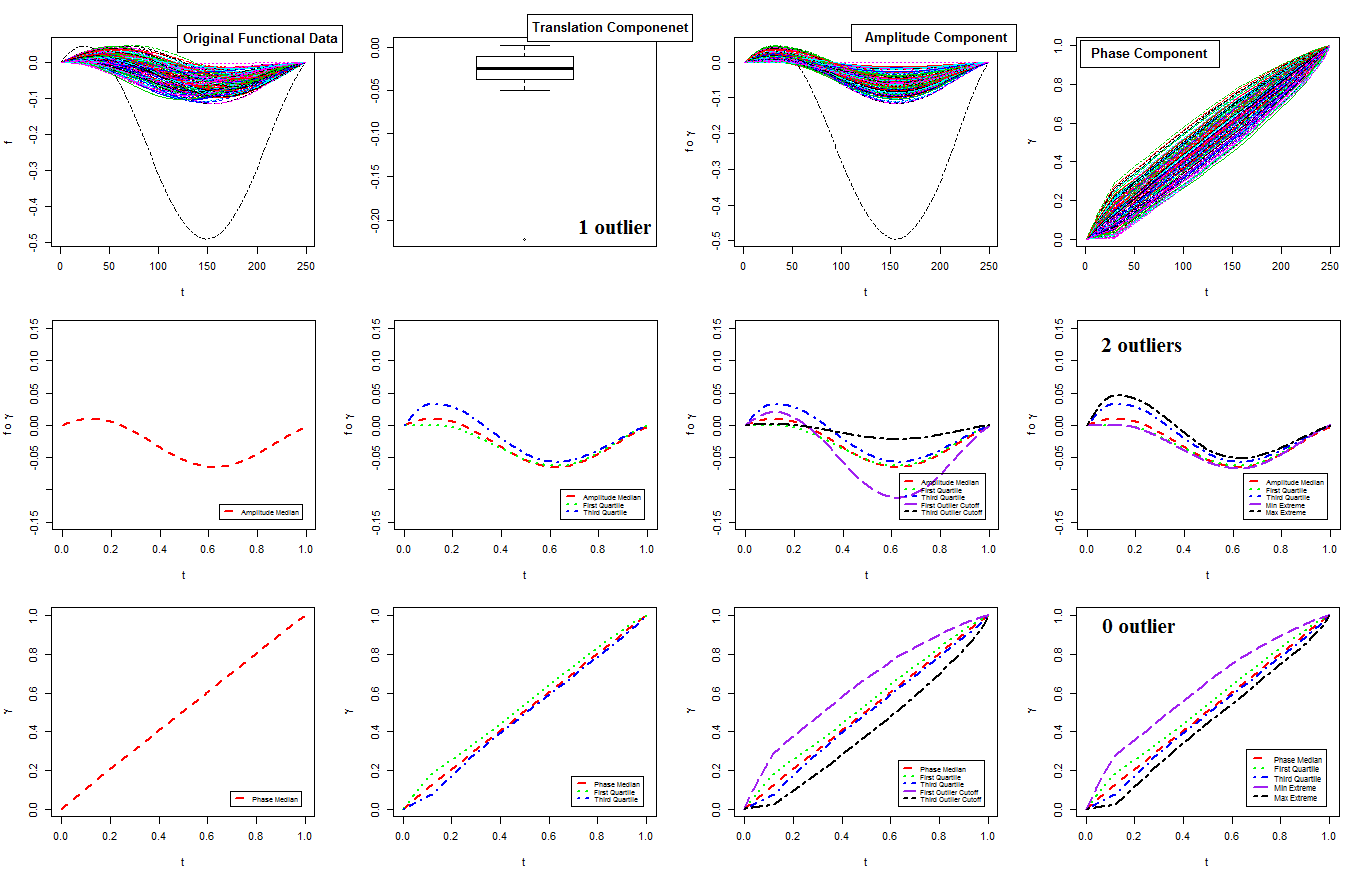}
	\end{center} \caption{Results of simulation 2 of benchmark geometries for $y(t)$.}\label{Shape3}
\end{figure}

\begin{figure}[!h] 
	\begin{center}
		\includegraphics[width=0.8\textwidth,angle =0, scale=1.22]{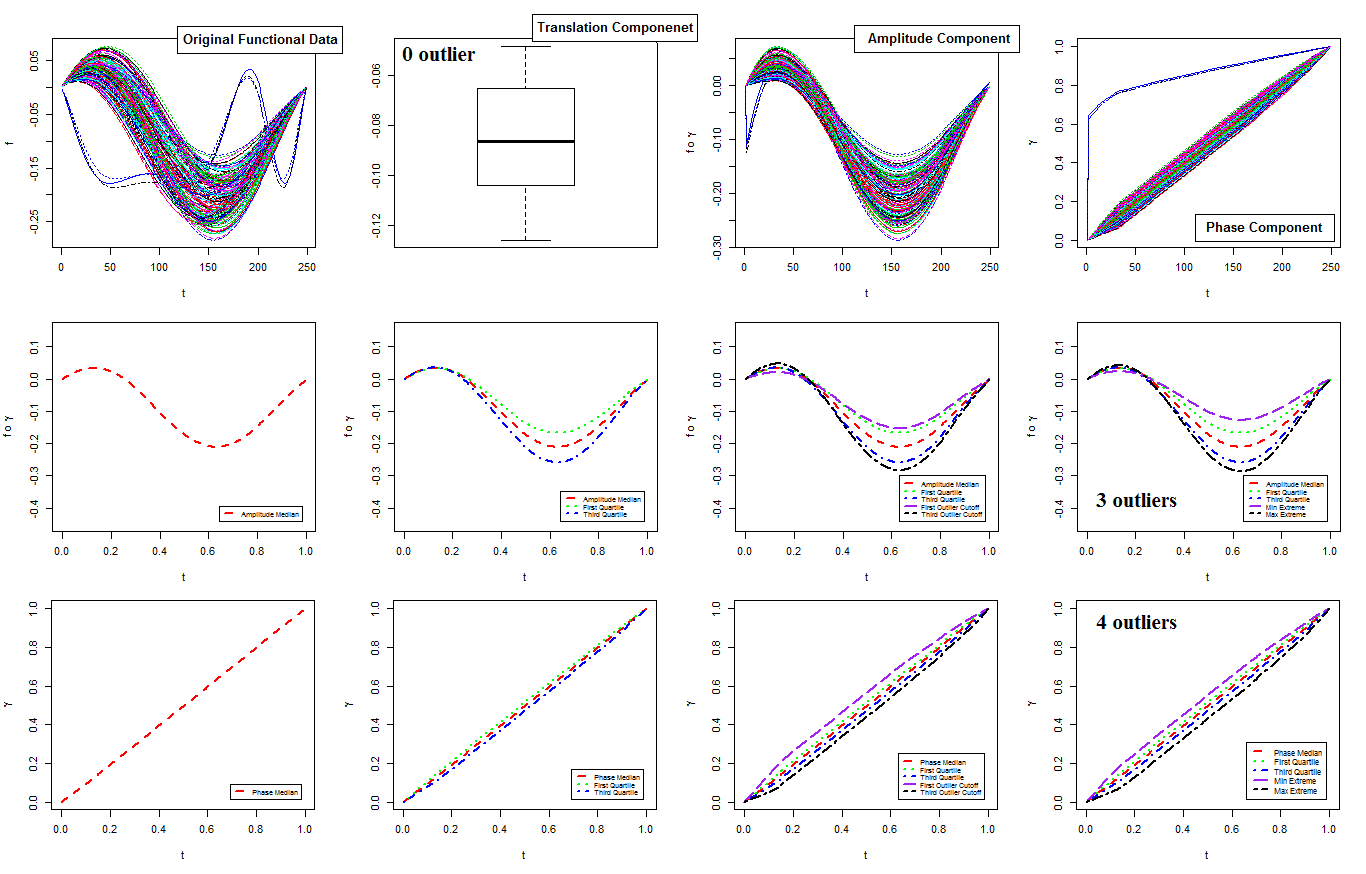}
	\end{center} \caption{Results of simulation 3 of benchmark geometries for $y(t)$.}\label{Shape4}
\end{figure}

\begin{figure}[!h] 
	\begin{center}
		\includegraphics[width=0.8\textwidth,angle =0, scale=1.2]{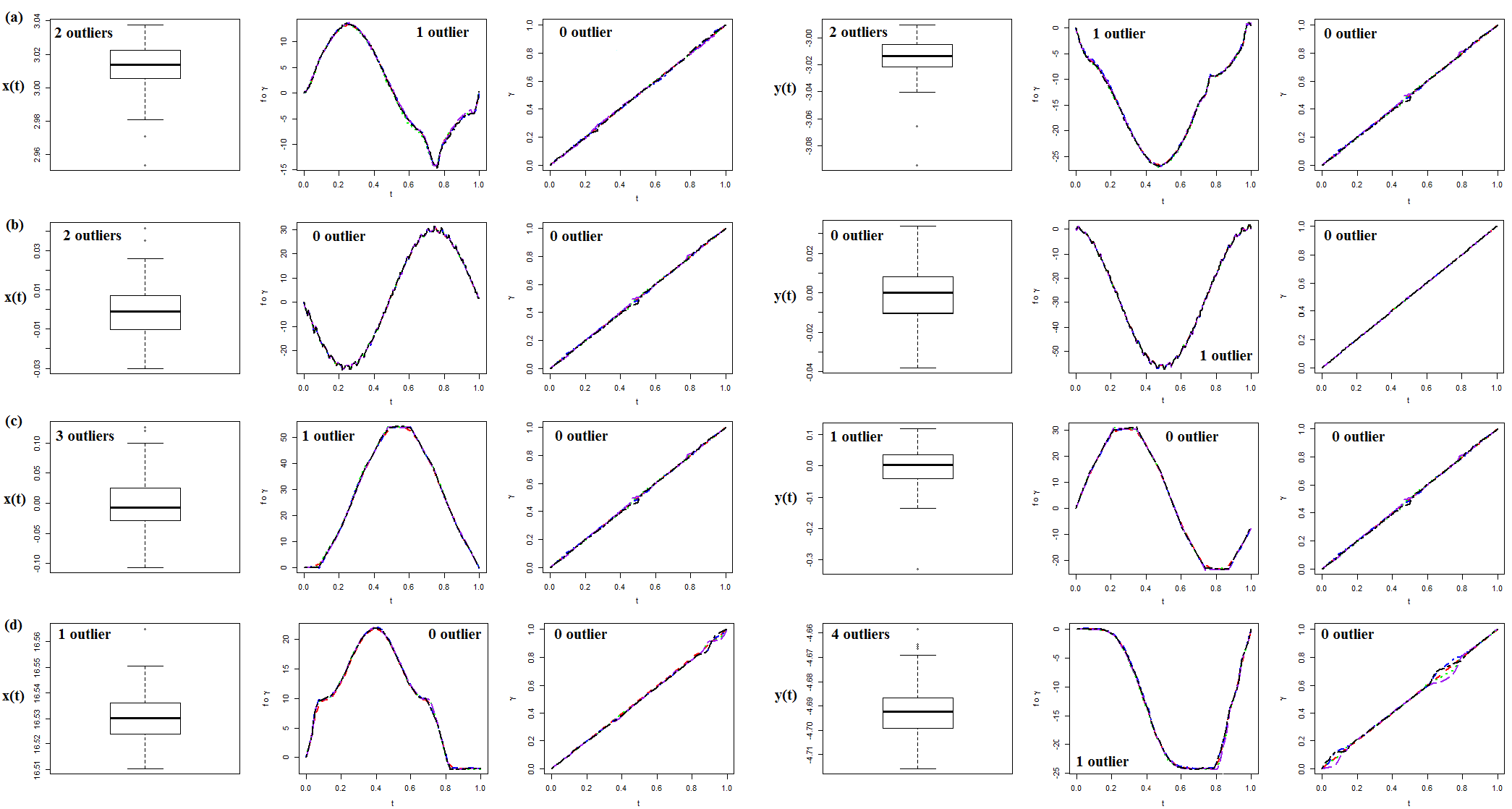}
	\end{center} \caption{Results of simulation 1 of Thingi10K geometries for (a) spiral bevel gear, (b) wheel, (c) logo, and (d) coin loading tube.}\label{Sim5PlotsThingi10K}
\end{figure}

\begin{figure}[!h] 
	\begin{center}
		\includegraphics[width=0.8\textwidth,angle =0, scale=1.2]{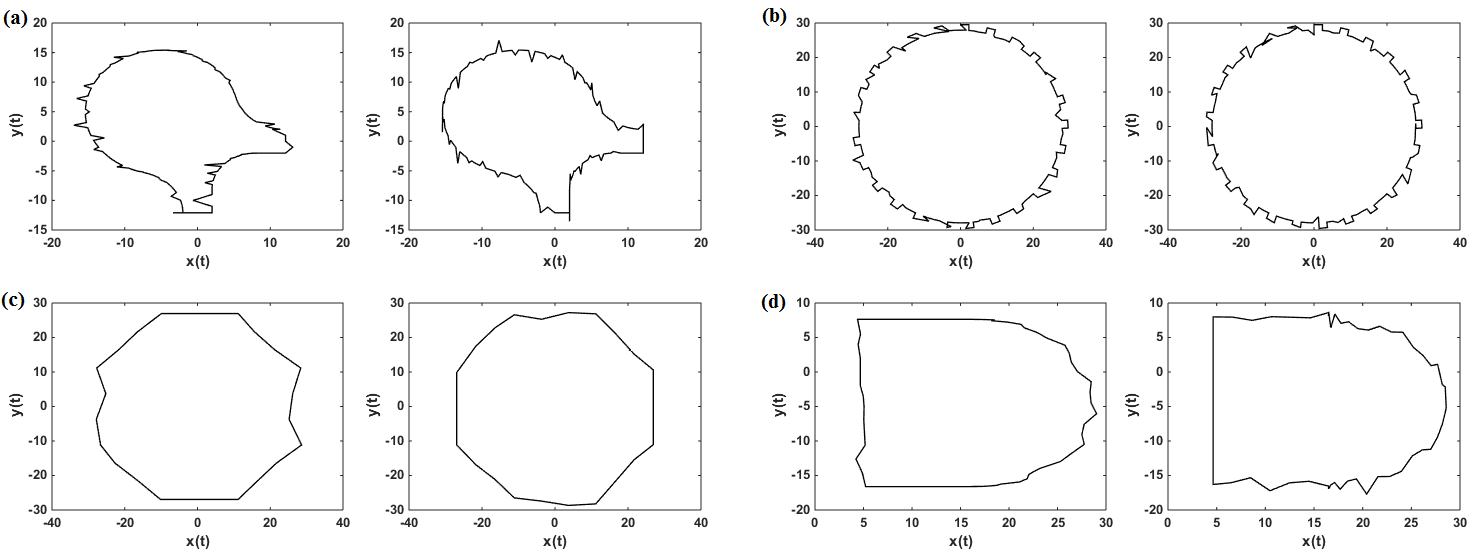}
	\end{center} \caption{Some of the detected contours for simulation 1 of Thingi10K geometries, (a) spiral bevel gear, (b) wheel, (c) logo, and (d) coin loading tube. The values of $x$ and $y$ are given in $mm$.}\label{OutlierSim5}
\end{figure}

\begin{figure}[!h] 
	\begin{center}
		\includegraphics[width=0.8\textwidth,angle =0, scale=1.2]{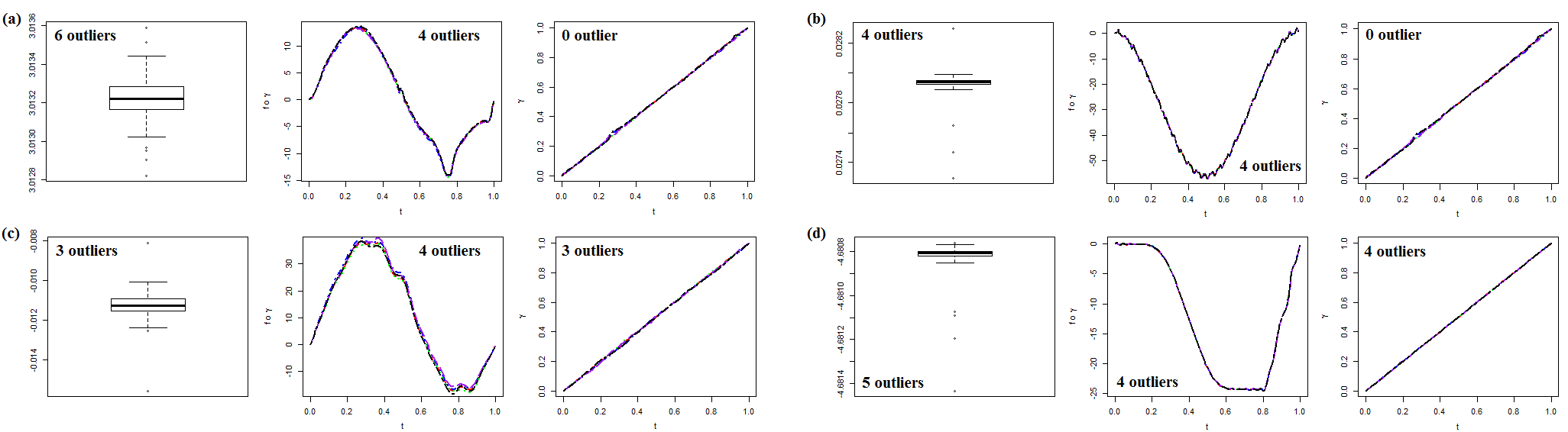}
	\end{center} \caption{Results of simulation 2 of Thingi10K geometries for (a) spiral bevel gear, (b) wheel, (c) logo, and (d) coin loading tube.}\label{Sim6PlotsThingi10K}
\end{figure}

\begin{figure}[!h] 
	\begin{center}
		\includegraphics[width=0.8\textwidth,angle =0, scale=1.2]{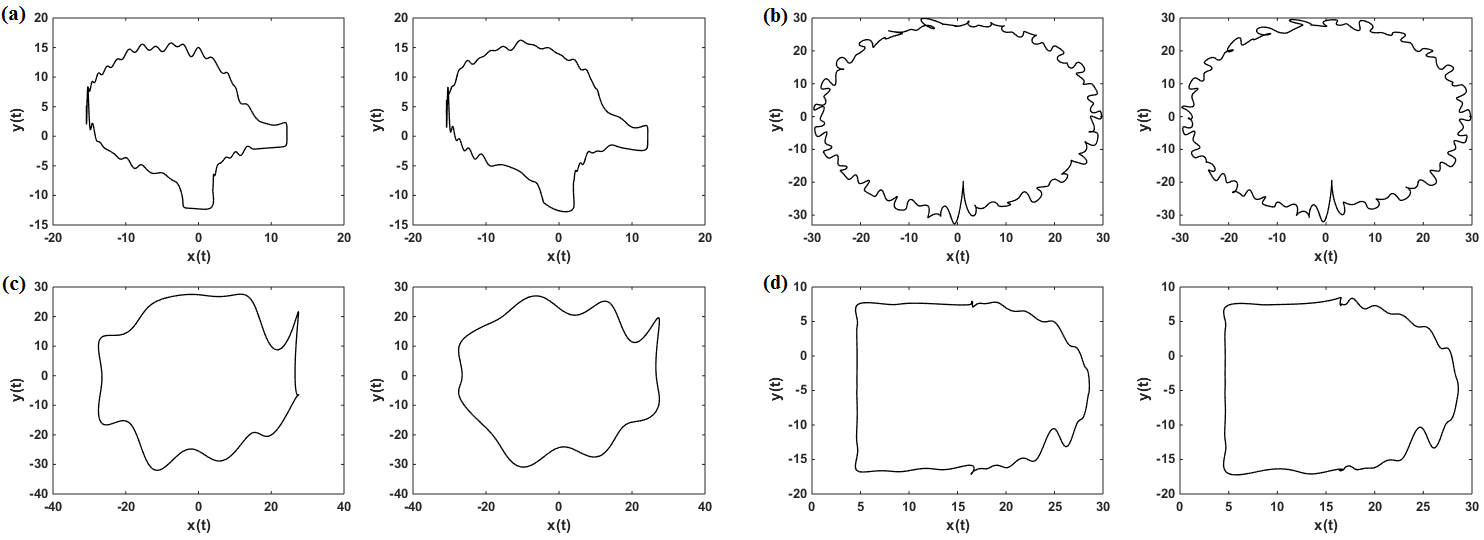}
	\end{center} \caption{Some of the detected contours for simulation 2 of Thingi10K geometries, (a) spiral bevel gear, (b) wheel, (c) logo, and (d) coin loading tube. The values of $x$ and $y$ are given in $mm$.}\label{OutlierSim6}
\end{figure}

\begin{figure}[!h] 
	\begin{center}
		\includegraphics[width=0.8\textwidth,angle =0, scale=1.2]{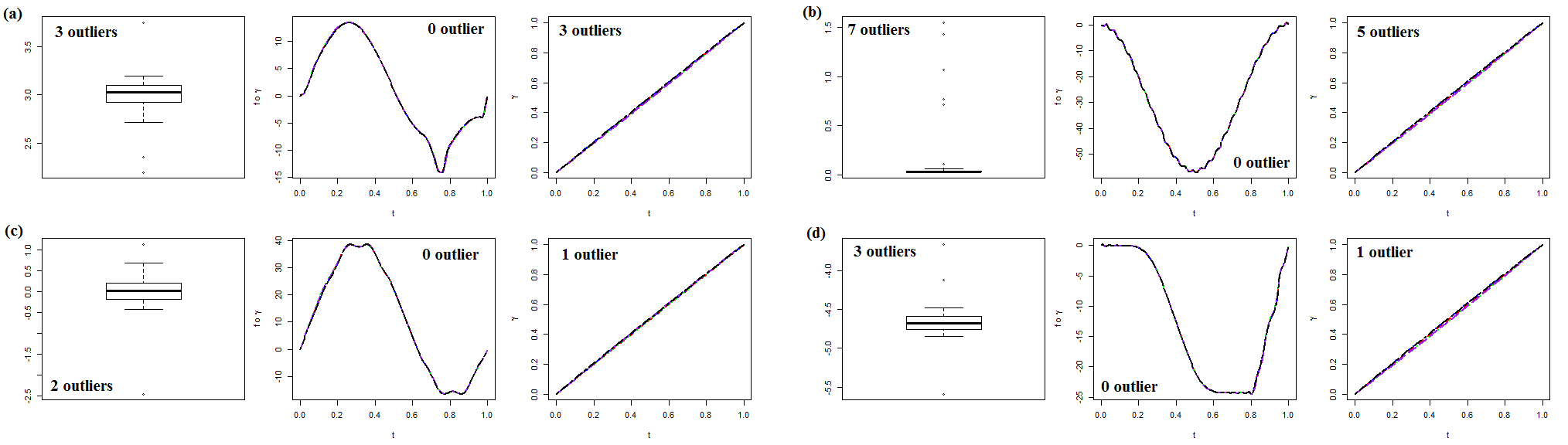}
	\end{center} \caption{Results of simulation 3 of Thingi10K geometries for (a) spiral bevel gear, (b) wheel, (c) logo, and (d) coin loading tube.}\label{Sim7PlotsThingi10K}
\end{figure}

\begin{figure}[!h] 
	\begin{center}
		\includegraphics[width=0.8\textwidth,angle =0, scale=1.2]{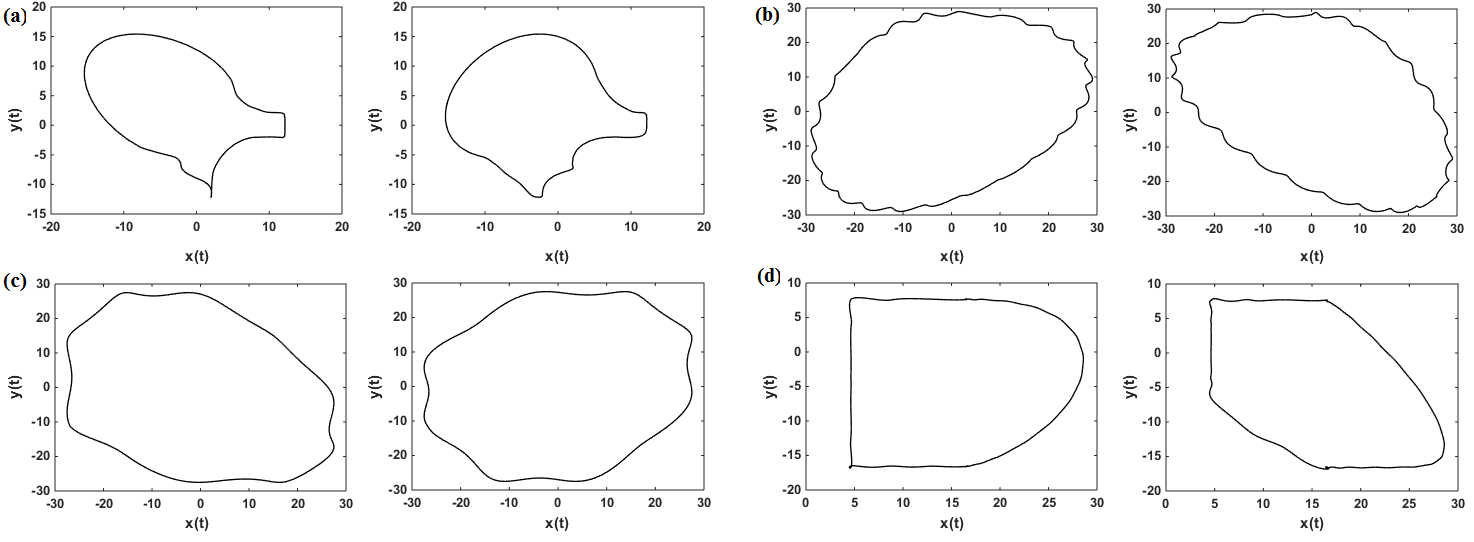}
	\end{center} \caption{Some of the detected contours for simulation 3 of Thingi10K geometries, (a) spiral bevel gear, (b) wheel, (c) logo, and (d) coin loading tube. The values of $x$ and $y$ are given in $mm$.}\label{OutlierSim7}
\end{figure}

\end{document}